\documentclass[12pt]{article}

\usepackage{tikz}
\usetikzlibrary{matrix,arrows}

\usepackage{a4wide}

\usepackage{amsmath,amssymb,amsfonts} 
\usepackage[T1]{fontenc}
\usepackage{yfonts}

\usepackage{extarrows}
\usepackage{undertilde}

\usepackage{mathrsfs,latexsym}

\newcommand{\II}{{\boldsymbol{1}}}

\newcommand{\CC}{{\mathbb C}}
\newcommand{\Kbb}{{\mathbb K}}
\newcommand{\RR}{{\mathbb R}}

\newcommand{\NN}{{\mathbb N}}

\newcommand{\ZZ}{{\mathbb Z}}

\newcommand{\CoinX}[1]{C_0^\infty({#1})}

\newtheorem{Thm}{Theorem}[section]

\newtheorem{Def}[Thm]{Definition}
\newtheorem{Lem}[Thm]{Lemma}
\newtheorem{Prop}[Thm]{Proposition}
\newtheorem{Cor}[Thm]{Corollary}

\numberwithin{equation}{section}

\newcommand{\Js}{{\mathsf J}}

\newcommand{\KK}{{\mathscr K}}

\newcommand{\Ac}{{\mathcal A}}
\newcommand{\Bc}{{\mathcal B}}

\newcommand{\Ic}{{\mathcal I}}

\newcommand{\OO}{{\mathscr O}}

\newcommand{\fb}{{\boldsymbol{f}}}
\newcommand{\gb}{{\boldsymbol{g}}}
\newcommand{\hb}{{\boldsymbol{h}}}
\newcommand{\nb}{{\boldsymbol{n}}}

\newcommand{\Tb}{{\boldsymbol{T}}}

\newcommand{\ogth}{{\mathfrak o}}
\newcommand{\tgth}{{\mathfrak t}}

\newcommand{\Ran}{{\rm Ran}\,}
\newcommand{\supp}{{\rm supp}\,}
\newcommand{\Span}{{\rm span}\,}

\newcommand{\im}{{\rm im}\,}

\newcommand{\Sym}{{\rm Sym}}

\newcommand{\dvol}{d\textrm{vol}}

\newcommand{\Ob}{{\boldsymbol{0}}}

\newcommand{\Bb}{{\boldsymbol{B}}}
\newcommand{\Cb}{{\boldsymbol{C}}}

\newcommand{\Lb}{{\boldsymbol{L}}}
\newcommand{\Mb}{{\boldsymbol{M}}}
\newcommand{\Nb}{{\boldsymbol{N}}}

\newcommand{\Mc}{{\mathcal{M}}}

\newcommand{\LCTo}{{\sf LCT}_0}
\newcommand{\LCT}{{\sf LCT}}
\newcommand{\Loco}{{\sf Loc}_0}
\newcommand{\Loc}{{\sf Loc}}
\newcommand{\Man}{\Loco}
\newcommand{\Mand}{\Loc}

\newcommand{\preSympl}{{\sf preSympl}}

\newcommand{\Sympl}{{\sf Sympl}}

\newcommand{\Alg}{{\sf Alg}}
\newcommand{\CAlg}{{\sf C^*\hbox{-}Alg}}

\newcommand{\Vect}{{\sf Vect}}
\newcommand{\iVect}{{\sf iVect}}
\newcommand{\Phys}{{\sf Phys}}

\newcommand{\Af}{{\mathscr A}}

\newcommand{\Bf}{{\mathscr B}}

\newcommand{\Cf}{{\mathscr C}}
\newcommand{\Df}{{\mathscr D}}

\newcommand{\Ff}{{\mathscr F}}

\newcommand{\Jf}{{\mathscr J}}

\newcommand{\Qf}{{\mathscr Q}}
\newcommand{\Rf}{{\mathscr R}}

\newcommand{\Tf}{{\mathscr T}}

\newcommand{\Wf}{{\mathscr W}}
\newcommand{\Zf}{{\mathscr Z}}

\newcommand{\Sol}{{\mathscr L}}

\newcommand{\id}{{\rm id}}

\newcommand{\nto}{\stackrel{.}{\to}}

\newcommand{\Aut}{{\rm Aut}}

\DeclareMathOperator{\cl}{cl}

\DeclareMathOperator{\eq}{eq}

\newcommand{\rce}{{\rm rce}}

\newcommand{\dyn}{{\rm dyn}}
\newcommand{\kin}{{\rm kin}}

\DeclareMathOperator{\CCR}{CCR}
\newcommand{\WW}{\mathcal{W}}

\begin{document}


\renewcommand{\thefootnote}{\fnsymbol{footnote}}
\phantom{*}
\vspace{-1in}

\noindent ESI Preprint 2336\\[0.5in]
\begin{center}
{ \Large \bf Dynamical locality of the free scalar field}
\\[20pt]
{\large  Christopher J.\ Fewster${}^{1}$\footnote{E-mail: chris.fewster@york.ac.uk}
 {\rm and}   Rainer Verch${}^{2}$\footnote{E-mail: verch@itp.uni-leipzig.de}}
\\[14pt]  
                 ${}^1$ Department of Mathematics,
                 University of York, 
                 Heslington,
                 York YO10 5DD, U.K.
                 \\
                 ${}^2$\,
Institut f\"ur Theoretische Physik,
Universit\"at Leipzig,
04009 Leipzig, Germany \\[14pt]
\today
\end{center}
${}$\\
{\small {\bf Abstract.}    
Dynamical locality is a condition on a locally covariant physical theory, asserting that
kinematic and dynamical notions of local physics agree. This condition was introduced
in [arXiv:1106.4785], where it was shown to be closely related to the question of  
what it means for a theory to describe the same physics on different spacetimes. 
In this paper, we consider in detail the example of the free 
minimally coupled Klein--Gordon field, both as 
a classical and quantum theory (using both the Weyl algebra and a smeared field approach). 
It is shown that the massive theory obeys dynamical locality, both classically and in quantum field theory, in all spacetime dimensions $n\ge 2$ and allowing for spacetimes with finitely many connected components. In contrast, the massless theory is shown to violate dynamical locality in any spacetime dimension, in both classical and quantum theory, owing to a rigid gauge symmetry. Taking this into account (equivalently, working with the massless current)
dynamical locality is restored in all dimensions $n\ge 2$ on connected spacetimes, and in all dimensions
$n\ge 3$ if disconnected spacetimes are permitted.  The results on the quantized theories are obtained
using general results giving conditions under which dynamically local classical symplectic theories have 
dynamically local quantizations.}
${}$

\renewcommand{\thefootnote}{\arabic{footnote}}

\section{Introduction}

In the functorial description of locally covariant quantum field theory in curved spacetimes introduced by  Brunetti, Fredenhagen and Verch
(BFV)~\cite{BrFrVe03}, a physical theory is described by a 
functor from a category of globally hyperbolic spacetimes to a category of $(C)^*$-algebras. This
view-point has proved fruitful in various aspects of model-independent quantum field theory, e.g., the proof of the spin-statistics connection \cite{Verch01}, analogues of the Reeh-Schlieder theorem \cite{Sanders_ReehSchlieder, Dapp:2011},
superselection theory~\cite{Br&Ru05,BrunettiRuzzi_topsect}, and the perturbative construction of interacting theories in curved spacetime \cite{BrFr2000,Ho&Wa01,Ho&Wa02}. Moreover, it has led to applications in cosmology~\cite{DapFrePin2008,DegVer2010,VerchRegensburg} and constraints on Casimir energy densities in cavities of arbitrary geometry~\cite{Few&Pfen06, Fewster2007}.

The same formalism can be applied to branches of physics other than quantum field theory by suitable choice of the target category. As we have emphasized in~\cite{FewVer:dynloc_theory} (see also~\cite{FewsterRegensburg}) the functorial framework allows us to analyse theories and relationships between theories at the level of the functors, rather than in individual spacetimes. In particular, 
we showed in~\cite{FewVer:dynloc_theory} how one can even address questions such as whether
a given theory can be regarded as representing the same physics in all spacetimes (SPASs). 
In particular, we gave a simple condition that should be obeyed by any reasonable notion of SPASs
and showed that (a) it is not satisfied by the full class of locally covariant theories, but (b) it is
satisfied by the subclass of theories obeying {\em dynamical locality}, which is free of
the known pathologies associated with the full class. 

The precise
definition of dynamical locality will be reviewed in Sect.~\ref{sect:review}; roughly, it requires that
kinematical and dynamical notions of localisation coincide for the theory. A variety of
properties of dynamically local theories were studied in~\cite{FewVer:dynloc_theory},
suggesting that the property is of independent interest in locally covariant physics.
For instance, it permits a general proof of the impossibility of selecting a 
single `natural' state in each spacetime for any nontrivial dynamically local theory that
reduces to a Haag--Kastler or Wightman theory in Minkowski space, with the supposed natural state as the vacuum (see~\cite[\S 6.3]{FewVer:dynloc_theory}). Previous results on this question were 
confined to free models, and even there were not absolutely complete arguments.  

The present paper provides examples to illustrate the theory developed in~\cite{FewVer:dynloc_theory}.
In particular, we will show that the massive minimally coupled Klein--Gordon theory obeys the
dynamical locality condition, but that the massless minimally coupled theory does not, unless
it is formulated as a theory of currents. This
property holds for both the classical and quantized theories, with quantization performed
either using a $*$-algebra of smeared fields, or the Weyl $C^*$-algebra approach.
Our results on the quantized theories are established as special cases of general results, 
which we prove, showing how dynamically local classical theories (valued in a category of symplectic spaces) have 
dynamically local quantizations under mild additional hypotheses. 

The failure of dynamical locality in the massless case may be traced to the existence of the rigid gauge freedom $\varphi\mapsto\varphi+\text{constant}$, which is normally ignored in treating this theory. Taking the gauge symmetry
seriously, and quantizing the theory on the same lines as electromagnetism and related theories~\cite{Dimock92, Few&Pfen03,Pfenning09}, we find that dynamical locality is restored with the sole exception of the theory in two-dimensional
(potentially disconnected) spacetimes. Dynamical locality holds if one restricts to connected spacetimes.
The significance of this sole exception is unclear and will provide the basis for further work.
In a separate work~\cite{Ferg_in_prep} it will be shown that the nonminimally coupled Klein--Gordon theory obeys dynamical locality for any value of the mass (in this case there is no gauge freedom). 
Studies of other theories are under way.

The paper is structured as follows. We review the basic ideas and terminology of~\cite{FewVer:dynloc_theory} in Sect.~\ref{sect:review}, and then discuss the dynamical
locality of the classical minimally coupled field and massless current in Sects.~\ref{sect:Klein_Gordon} and~\ref{sect:currents}. Our general results on quantization by the infinitesimal Weyl algebra and Weyl algebra approaches appear in Sects.~\ref{sect:infinitesimal} and~\ref{sect:Weyl}.  Remarks on related approaches appear in Sect.~\ref{sect:outlook}. The appendices give background on some multilinear algebra required in the body of the paper, and also establish the differentiability of the relative Cauchy evolution for the real scalar field. 

\section{Dynamical locality}\label{sect:review}

We briefly summarise the BFV approach to locally covariant physics, as elaborated in~\cite{FewVer:dynloc_theory}. This is a framework for studying physics on 
fixed spacetimes, which are taken to be globally hyperbolic, but not necessarily connected. 
The fundamental definitions of category theory~\cite{MacLane,AdamekHerrlichStrecker} will be assumed, but some particular structures will be defined where necessary. 

\paragraph{Spacetimes}

A spacetime of dimension $n\ge 2 $ is a quadruple $(\Mc,\gb,\ogth,\tgth)$ such that $\Mc$ is a smooth, paracompact, orientable nonempty 
$n$-manifold with finitely many connected components, $\gb$ is a smooth time-orientable metric of signature $+-\cdots-$ on $\Mc$, and $\ogth$ and $\tgth$ are choices of orientation and time-orientation
respectively. A spacetime is said to be {\em globally hyperbolic} if it contains no closed causal curves and the intersection of the  causal past and future of any pair of points is
compact. It is sufficient that $\Mc$ contains
a Cauchy surface, i.e., a subset met exactly once by every inextendible timelike curve in the spacetime.
(A number of properties of globally hyperbolic spacetimes, and comments on different notions appearing in the literature may be found in Sect.~2 of~~\cite{FewVer:dynloc_theory}.) 

The globally hyperbolic spacetimes (of dimension $n$) form the
objects of a category $\Mand$. By definition, a morphism $\psi$ in $\Mand$ between
$\Mb=(\Mc,\gb,\ogth,\tgth)$ and $\Mb'=(\Mc',\gb',\ogth',\tgth')$ is a smooth embedding (also denoted
$\psi$) of $\Mc$ in $\Mc'$ whose image is causally convex\footnote{That is, it
contains all causal curves of which it contains the endpoints.}
 in $\Mb'$ and
such that $\psi^*\gb'=\gb$, $\psi^*\ogth'=\ogth$ and $\psi^*\tgth'=\tgth$. 
Thus the embedding is isometric and respects orientation and
time-orientation. The full subcategory of $\Mand$ with connected spacetimes as objects will be denoted $\Man$. 

Two particular classes of $\Mand$ and $\Man$ morphisms will be used extensively in what follows: {\em canonical inclusions} and {\em Cauchy morphisms}. Inclusions arise as follows. 
For any $\Mb$ in $\Mand$ (and hence $\Man$) let $\OO(\Mb)$ be the set of 
open globally hyperbolic subsets of $\Mb$ with at most finitely many connected components all of which are mutually causally disjoint, and let $\OO_0(\Mb)$ be the set of connected open globally hyperbolic subsets of $\Mb$. 
For each $\Mb=(\Mc,\gb,\ogth,\tgth)\in\Mand$, any nonempty $O\in \OO(\Mb)$ induces an object 
$\Mb|_O=(O,\gb|_O,\ogth|_O,\tgth|_O)$ of $\Mand$, which
we call the {\em restriction} of $\Mb$ to $O$, and the subset inclusion of $O$ in $\Mb$
induces a $\Mand$-morphism $\iota_{\Mb;O}:\Mb|_O\to\Mb$ that we call a canonical inclusion. 
If $O\in\OO_0(\Mb)$ for $\Mb\in\Man$ then
$\iota_{\Mb;O}$ is also a $\Man$-morphism, provided $O$ is nonempty. A {\em Cauchy morphism}
is a morphism $\psi:\Mb\to\Nb$ whose image $\psi(\Mb)$ contains a Cauchy surface for $\Nb$. 

\paragraph{Physical systems}

The categories $\Man$ and $\Mand$ provide the arena for locally covariant physics. 
Physical systems themselves are described as objects in a category $\Phys$, which is 
determined by the type of physical system under consideration. The general conditions
imposed on $\Phys$ in~\cite{FewVer:dynloc_theory} are: 
\begin{itemize}\addtolength{\itemsep}{-0.5\baselineskip}
\item all morphisms in $\Phys$ are monic, i.e., $f\circ g = f\circ h$ implies $g=h$ for
all $f,g,h$;
\item there is an initial object, $\Ic$, i.e., for every object $A$ of $\Phys$, there is a 
exactly one morphism from $\Ic$ to $A$, denoted $\Ic_A$; 
\item it has equalizers, i.e., for any pair of morphisms $f,g:A\to B$ there is a morphism $h$ 
(the equalizer of $f$ and $g$)  such that $f\circ h= g \circ h$ and such that
if $k$ is any morphism such that $f\circ k=g\circ k$ then $k$ factorizes
uniquely via $h$, i.e., $k=h\circ m$ for a unique morphism $m$ ---
we write $h\cong \eq(f,g)$;
\item it has intersections and unions (see~\cite{DikranjanTholen}, and Appendix~B to~\cite{FewVer:dynloc_theory}).
\end{itemize}
Here, intersections and unions are not generally identical to the intersections and unions of set theory, but appropriate generalizations to the category in question, e.g., the `union' in a category of 
vector spaces is determined by the linear span etc. Like equalizers, they are defined by universal properties, only up to isomorphism: if $f:A\to B$ is an intersection of the morphisms $f_i:A_i\to B$, and $g:A'\to A$ is an isomorphism, then $f\circ g:A'\to B$ is also an intersection of the $f_i$. We write 
$f\cong \bigwedge_i f_i$ in such circumstances, and likewise denote a union by 
$f\cong\bigvee_i f_i$. In concrete categories where the morphisms can be regarded as functions (with particular structure) and composition is composition of the functions, the `up to isomorphism' nature
of these definitions can largely be ignored, because the image of such a map is unchanged by 
precomposition with an isomorphism.

To illustrate the definitions, we give the examples which will be relevant to us in the present paper. In the quantum theory, we will use $\Alg$, the category of unital complex $*$-algebras, with unit-preserving injective $*$-homomorphisms as the morphisms, and $\CAlg$, the full subcategory of $\Alg$ consisting of $C^*$-algebras. The initial object is the algebra of complex numbers with $1$ as unit
and complex conjugation as the $*$-operation and (for $\CAlg$) the complex
modulus as $C^*$-norm. Subobjects may be identified with $(C)*$-subalgebras; 
the intersection is the ordinary intersection of $(C)*$-subalgebras and the union is the $(C)*$-algebraic span;
the equalizer of $\alpha,\beta:\Ac\to\Bc$ can be described as the inclusion map in $\Ac$ of the maximal $(C)*$-subalgebra of $\Ac$ on which $\alpha$ and $\beta$ agree. 

For the classical theory, we will consider various categories of
(pre)symplectic spaces. Let $\Kbb$ be either $\RR$ or $\CC$. A {\em pre-symplectic space} over $\Kbb$ is a pair $(V,\sigma)$ consisting of a $\Kbb$-vector space $V$ equipped with an antisymmetric $\Kbb$-bilinear\footnote{{\em Not} sesquilinear, in the complex case.} form $\sigma:V\times V\to \Kbb$. In the special case where, to each  nonzero $u\in V$, there is a $v\in V$ with $\sigma(u,v)\neq 0$, we say that
$(V,\sigma)$ is a weakly nondegenerate symplectic space. A symplectic map between two pre-symplectic spaces $(V,\sigma)$ and $(V',\sigma')$ is a $\Kbb$-linear map $f:V\to V'$ such that $\sigma'(f u,fv) = \sigma(u,v)$ for all $u,v\in V$. We define a category
$\preSympl_\RR$ to be the category of real pre-symplectic spaces, with injective symplectic maps as morphisms. 

In the complex case, we wish to have available a complex conjugation as well. 
Thus $\preSympl_\CC$ will denote the category whose objects are triples $(V,\sigma,C)$, 
where $(V,\sigma)$ is a complex pre-symplectic space and $C:V\to V$ is an antilinear involution with
$\sigma(C u,C v) = \overline{\sigma(u,v)}$, and with morphisms $(V,\sigma,C)\to (V',\sigma',C')$
given by an injective symplectic maps $f:(V,\sigma)\to (V',\sigma')$ 
such that $C'\circ f=  f\circ C$. 
We may regard $\preSympl_\CC$ as the category of complexified real pre-symplectic spaces. The full subcategory of $\preSympl_\Kbb$ whose objects are weakly nondegenerate will be denoted $\Sympl_\Kbb$. (Note that if $(V,\sigma)$ is weakly nondegenerate then any symplectic map with domain $(V,\sigma)$ is injective.)

Given any $(V,\sigma)\in\preSympl_\RR$  any vector subspace $W$ of $V$ induces $(W,\sigma|_{W\times W})\in\preSympl_\RR$ and a canonical inclusion morphism $(W,\sigma|_{W\times W})\to (V,\sigma)$ in $\preSympl_\RR$. Similarly, given $(V,\sigma,C)\in \preSympl_\CC$, 
any $C$-invariant subspace $W$ of $V$ induces analogous structures in $\preSympl_\CC$. 
(Weak nondegeneracy is not necessarily inherited under this restriction, which is why we work with pre-symplectic spaces.)
The trivial vector space with zero symplectic form\footnote{This is non-degenerate, despite initial appearances.}
(and trivial complex conjugation if $\Kbb=\CC$) provides an initial object in $\preSympl_\Kbb$. Intersections and unions are obtained in an obvious way from intersections and spans of ($C$-invariant) vector subspaces; 
the equalizer of $f,g:(V,\sigma)\to (V',\sigma')$ in $\preSympl_\RR$ is the canonical inclusion morphism induced by the subspace $\ker (f-g)$ of $V$; in the complex case, with 
$f,g:(V,\sigma,C)\to (V',\sigma',C')$, the subspace $\ker (f-g)$ is $C$-invariant, and again defines 
the equalizer. Accordingly, $\preSympl_\RR$ and $\preSympl_\CC$ meet our general conditions to be categories of physical systems. 

There is a useful functor $\Rf:\preSympl_\CC\to \preSympl_\RR$:
to each $(V,\sigma,C)$, it assigns $(V^C,\sigma|_{V^C\times V^C})$, where $V^C:=\ker (C-\id_V)$, regarded as a real vector space; to any morphism $f:(V,\sigma,C)\to (V',\sigma',C')$, it
assigns the restriction $f|_{V^C}$, whose range is easily seen to lie in $W^{C'}$. Moreover, 
this functor preserves weak nondegeneracy, intersections, unions and equalizers: for $C$-invariant subspaces $U,U'$ of $V$,
we have $(U\cap U')^C = U^C\cap U^{\prime C}$, $(U +  U')^C = U^C + U^{\prime C}$, while
for morphisms $f,g$, $(\ker (f-g))^C = \ker (f|_{V^C}-g|_{V^C})$.

\paragraph{Locally covariant physical theories}

A locally covariant physical theory assigns physical systems to spacetimes and, importantly, to each
hyperbolic embedding of spacetimes it assigns an embedding of the corresponding physical systems. 
It is represented by a (covariant) functor $\Af:\Mand\to\Phys$ (or from $\Man$ if one restricts to connected spacetimes). The theories form the objects of a category $\LCT$ (or $\LCTo$ for theories on $\Man$) in
which morphisms between theories are natural transformations between the corresponding functors; this
was used intensively in~\cite{FewVer:dynloc_theory} but will not be needed here.

An important general feature of the BFV framework is that it contains a natural notion of dynamics:
{\em relative Cauchy evolution}. Let $\Mb=(\Mc,\gb,\ogth,\tgth)\in\Mand$ be a globally hyperbolic spacetime. Given any symmetric $\hb\in\CoinX{T^0_2\Mb}$ such that $\gb+\hb$ is a time-orientable Lorentz metric on $\Mc$, there is a
unique choice of time-orientation $\tgth_\hb$ for $\gb+\hb$ that agrees with $\tgth$ outside
$K$. If $\Mb[\hb]=(\Mc,\gb+\hb,\ogth,\tgth_\hb)$ is a globally hyperbolic
spacetime, we say that $\hb$ is a {\em globally hyperbolic perturbation}
of $\Mb$ and write $\hb\in H(\Mb)$. The subset of $\hb\in H(\Mb)$ with
support in $K\subset\Mc$ is denoted $H(\Mb;K)$. Clearly, $\Mb=\Mb[\Ob]$, 
where $\Ob$ is identically zero, and indeed $H(\Mb)$ contains an open neighbourhood of
$\Ob$ in the usual test-function topology on symmetric smooth
compactly supported sections of $T^0_2\Mb$ (see \S 7.1
of~\cite{BeemEhrlichEasley}). 
We endow $H(\Mb)$ with the subspace topology induced from $\Df(T^0_2M)$.

For each $\hb\in H(\Mb)$, set $\Mc^\pm=\Mc\setminus J^\mp_\Mb(\supp\hb)$,
where $J^{+/-}_\Mb(S)$ denotes the causal future/past of a set $S$ (see~\cite{ONeill} for
definitions relating to causal structure). As shown in
Sect.~3.4 of~\cite{FewVer:dynloc_theory}, the $\Mc^\pm$ are globally hyperbolic
subsets of both $\Mb$ and $\Mb[\hb]$ and therefore define canonical inclusions $\imath_\Mb^\pm[\hb]:\Mb^\pm[\hb]\to \Mb$  and $\jmath_\Mb^\pm[\hb]\to\Mb[\hb]$, where 
$\Mb^\pm[\hb]:=\Mb|_{\Mc^\pm}$. Moreover, these canonical inclusions are Cauchy. 

Then for any theory satisfying  the {\em time-slice property}, which
requires that the functor should map each Cauchy morphism to an isomorphism in $\Phys$, 
we obtain isomorphisms
\[
\tau^\pm_\Mb[\hb]=\Af(\jmath_\Mb^\pm[\hb])
\circ(\Af(\imath_\Mb^\pm[\hb]))^{-1}:\Af(\Mb)\to\Af(\Mb[\hb])
\]
and an automorphism $\rce_\Mb[\hb]$ of $\Af(\Mb)$ given by
\[
\rce_\Mb[\hb] = (\tau^-_\Mb[\hb])^{-1} \circ \tau^+_\Mb[\hb],
\]
which is called the  relative Cauchy evolution induced by $\hb$. This formulation
of the relative Cauchy evolution is equivalent to that given in BFV, by~\cite[Prop.~3.3]{FewVer:dynloc_theory}.

\paragraph{Dynamical locality} 

BFV emphasised that the standard structures of algebraic quantum field theory can be recovered from the 
locally covariant approach, on specialisation to particular spacetimes. The same can be done in the general
case. Recalling that $\OO(\Mb)$ is the set of globally hyperbolic open subsets of $\Mb$
with at most finitely many mutually causally disjoint connected components, each 
nonempty $O\in\OO(\Mb)$ induces a  canonical inclusion
$\iota_{\Mb;O}:\Mb|_O\to\Mb$. Any locally covariant theory $\Af:\Mand\to\Phys$, 
then assigns a physical system $\Af(\Mb|_O)$ and a morphism $\Af(\iota_{\Mb;O}):\Af(\Mb|_O)\to\Af(\Mb)$ embedding this as a subsystem of the physical system assigned to $\Mb$.

Accordingly, let $\Phys$ be any category obeying our minimal assumptions and let 
$\Af\in\LCT$ (resp., $\LCTo$). For $\Mb\in\Mand$ (resp., $\Man$) and nonempty $O\in\OO(\Mb)$ (resp., $O\in\OO_0(\Mb)$), we define
\[
\Af^\kin(\Mb;O) = \Af(\Mb|_O), \qquad \text{and}\qquad
\alpha^\kin_{\Mb;O}=\Af(\iota_{\Mb;O}):\Af^\kin(\Mb;O)\to\Af(\Mb).
\]
We refer to the assignment $O\mapsto \alpha^{\kin}_{\Mb;O}$ as the {\em kinematic net}.

In general categories, it is better to focus on the morphism $\alpha^\kin_{\Mb;O}:= \Af(\iota_{\Mb;O})$, than its image in $\Af(\Mb)$, but for categories $\Phys$ such as those discussed in this paper, there is
little harm in identifying $\Af^\kin(\Mb;O)$ with this image. 

One of the main ideas in~\cite{FewVer:dynloc_theory} is that we may also use dynamics to identify local physics in theories obeying the time-slice property. This is done as follows. If $K$ is a compact subset of globally hyperbolic spacetime $\Mb$, 
any hyperbolic perturbation $\hb\in H(\Mb;K^\perp)$
represents a modification in the spacetime in regions causally inaccessible from $K$.
We may test the sensitivity of subsystems of $\Af(\Mb)$ to these metric perturbations using the
relative Cauchy evolution; in particular, we identify those subsystems that are insensitive to
all such perturbations in $H(\Mb;K^\perp)$ as candidates for being localised in $K$. 

In the cases $\Phys=\Alg,\CAlg$, this motivates the definition of a subalgebra
\[
\Af^{\bullet}(\Mb;K) = \{A\in\Af(\Mb): \rce_\Mb[\hb]A = A ~\textrm{for all
$\hb\in H(\Mb;K^\perp)$}\};
\]
more generally, we may define a morphism $\alpha^\bullet_{\Mb;K}$ as
the unique (up to isomorphism)
subobject of $\Af(\Mb)$ such that (i)
\begin{equation} \label{eq:alpha_bullet}
\rce_\Mb[\hb]\circ\alpha^\bullet_{\Mb;K} = \alpha^\bullet_{\Mb;K} \qquad
\forall \hb\in H(\Mb;K^\perp);
\end{equation}
and (ii) if any other morphism $\alpha$ satisfies
Eq.~\eqref{eq:alpha_bullet} in place of $\alpha^\bullet_{\Mb;K}$, then
there is a unique morphism $\beta$ with $\alpha =\alpha^\bullet_{\Mb;K}\circ\beta$.
The existence of $\alpha^\bullet_{\Mb;K}$ follows from the structural assumptions on $\Phys$ ---
indeed, we may write
\[
\alpha^\bullet_{\Mb;K}\cong \bigwedge_{\hb\in H(\Mb;K^\perp)} \eq(\rce_\Mb[\hb],\id_{\Af(\Mb)}),
\]
where $\bigwedge$ denotes the intersection in $\Phys$. 
To obtain the local physics on a general $O\in\OO(\Mb)$, we take the $\Phys$-union over a suitable class of
compact subsets of $O$,
\[
\alpha^\dyn_{\Mb;O} \cong \bigvee_{K\in\KK_b(\Mb;O)} \alpha^\bullet_{\Mb;K}.
\]
Here $K\in \KK_b(\Mb;O)$ if is a finite union of causally disjoint subsets of $O$, each of which is the
closure of a Cauchy ball $B$ with a relatively compact Cauchy development $D_\Mb(B)$; 
a Cauchy ball $B$ is a subset of a Cauchy surface, for which there is a chart containing the
closure of $B$, and in which $B$ is a nonempty open ball. (We also set $\KK_b(\Mb;\emptyset)=\{\emptyset\}$ by convention.)  This differs slightly from the 
definition given first in~\cite{FewVer:dynloc_theory}, but is equivalent by 
Lemma~5.3 of that reference.  As shown in~\cite{FewVer:dynloc_theory}, in spacetime dimension $n\ge 3$, each $K\in\KK_b(\Mb;O)$ has the property that $K^\perp$ has connected intersection with each connected component of $\Mb$; that this is not true in $n=2$ dimensions will have an interesting
consequence in Sect.~\ref{sect:currents}.

Given these definitions, dynamical locality is defined as follows: 
\begin{Def} \label{def:dynloc}
A theory $\Af\in \LCT$ (resp., $\LCTo$) obeys {\em dynamical locality} if it obeys
the timeslice property and, additionally, for each
$\Mb\in\Mand$ (resp., $\Man$) and all nonempty $O\in\OO(\Mb)$ (resp., $\OO_0(\Mb)$) we have $\Af^\kin(\Mb;O)\cong \Af^{\dyn}(\Mb;O)$, i.e., more abstractly, $\alpha^\kin_{\Mb;O} \cong \alpha^\dyn_{\Mb;O}$.
\end{Def} 
It is the main purpose of the present paper to investigate the extent to which this condition holds for the specific example of 
the minimally coupled Klein--Gordon field. 

Before proceeding, we note one abstract result that will be useful to us. Suppose
that $\Phys_1$ and $\Phys_2$ are two categories of physical systems meeting the general criteria above, and let $\Ff:\Phys_1\to\Phys_2$ be a functor that preserves  intersections, unions and equalizers, i.e.,
\[
\bigwedge_i \Ff(\alpha_i)\cong \Ff(\bigwedge_i\alpha_i),
\qquad \bigvee_i \Ff(\alpha_i)\cong \Ff(\bigvee_i\alpha_i), \qquad
\eq(\Ff(\alpha_1),\Ff(\alpha_2))\cong \Ff(\eq(\alpha_1,\alpha_2))
\]
for collections $(\alpha_i)$ of $\Phys_1$-morphisms. 
Then any physical theory $\Af_1:\Loc\to\Phys_1$ induces a theory $\Af_2=\Ff\circ\Af_1:\Loc\to\Phys_2$;
moreover, if $\Af_1$ is dynamically local, then so is $\Af_2$. In particular, any dynamically local
theory $\Sol:\Loc\to\preSympl_\CC$ induces a dynamically local theory $\Rf\circ\Sol:\Loc\to
\preSympl_\RR$.

\paragraph{The SPASs property} The motivation underlying~\cite{FewVer:dynloc_theory} is to understand the conditions under which a given theory may be regarded as displaying the same physics in all spacetimes. This is a difficult issue to make formal and it is conceivable that there could be a range of differing ways of doing so; however, we argued that (as well as local covariance) any such definition
should have the following property: if two theories are given, each of which [under the given definition] individually represents the same physics in all spacetimes, and one theory is a subtheory of the other, and they coincide in one particular spacetime, then they coincide in all spacetimes. In the functorial 
context, this is made precise when one understands `$\Af$ is a subtheory of $\Bf$' to mean that
there is a natural transformation $\eta:\Af\nto\Bf$, and `$\Af$ coincides with $\Bf$ in $\Mb$' to mean
that the component $\eta_\Mb$ is an isomorphism. We call this the {\em SPASs property}. Then the theories coincide in all spacetimes if and only if the functors $\Af$ and $\Bf$ are naturally isomorphic, which is the usual understanding of functorial equivalence. 

In~\cite{FewVer:dynloc_theory} it was shown that the class of all locally covariant theories (even restricting 
to those with the time-slice property) is far too large to have the SPASs property. Thus there is more 
to the issue of SPASs than simple local covariance. However, the dynamically local theories {\em do}
have the SPASs property (see Sect.~6.2 in~\cite{FewVer:dynloc_theory}).  While no claim is made that 
dynamical locality is the only possible definition that would qualify as a notion of SPASs, nor that the SPASs property is the only requirement one might reasonably demand of such a notion, it is currently the only contender. For this reason, and because dynamically local theories have a number of other nice properties, it is important to show that the main examples of quantum field theory in curved spacetime obey dynamical locality. This is the task to which we now turn.

\section{Classical theory of the Klein--Gordon field} \label{sect:Klein_Gordon}

\paragraph{Functorial definition}

Given any $\Mb\in \Mand$, the minimally coupled Klein--Gordon equation is 
\[
P_\Mb\phi:=(\Box_\Mb+m^2)\phi=0,
\]
where $m\ge 0$ is constant. We write $\Sol_\Kbb(\Mb)$ to denote the space of smooth 
$\Kbb$-valued solutions
to this equation that have compact support on Cauchy surfaces in
$\Mb$, and equip $\Sol_\Kbb(\Mb)$ with an antisymmetric bilinear form
\begin{equation}
\sigma_\Mb(\phi,\phi') = \int_\Sigma \left(\phi n^a\nabla_a\phi' - \phi'
n^a\nabla_a\phi\right) d\Sigma,
\end{equation}
where $\Sigma$ is any Cauchy surface with future-pointing unit normal $n^a$; its values are independent of the choice of $\Sigma$. In the case where $\Kbb=\CC$, we use complex conjugation of functions
as the antilinear involution on $\Sol_\CC(\Mb)$, i.e., $C_\Mb\phi=\bar{\phi}$. It is clear
that $(\Sol_\RR(\Mb),\sigma_\Mb)\in\preSympl_\RR$ and $(\Sol_\CC(\Mb),\sigma_\Mb,C_\Mb)\in\preSympl_\CC$. We will focus on the complex case, from which we may
read off all the structure of the real case by applying the functor $\Rf:\preSympl_\CC\to\preSympl_\RR$. To unburden the notation, we write $\Sol_\CC$ as $\Sol$
for the rest of this section. 

A number of standard facts will be used in the sequel, and are collected here for
reference. Let $E^\pm_\Mb:\CoinX{\Mb}\to C^\infty(\Mb)$ be the advanced ($-$) and retarded ($+$)
fundamental solutions, such that $\supp E^\pm_\Mb f\subset J^\pm_\Mb(\supp f)$. Then the advanced-minus-retarded
fundamental solution is $E_\Mb=E^-_\Mb-E^+_\Mb$, and acts as a bilinear form on $\CoinX{\Mb}$ by
\begin{equation}
E_\Mb(f,f')=(E_\Mb f')(f) \qquad (f,f'\in\CoinX{\Mb}).
\end{equation}
Then (see, e.g., \cite{BarGinouxPfaffle} for proofs) $E_\Mb$ has range and kernel
\[
\Sol(\Mb)=E_\Mb\CoinX{\Mb}, \qquad \ker E_\Mb=P_\Mb\CoinX{\Mb}
\]
and there is an identity
\begin{equation}\label{eq:sigma_and_E_1}
\sigma_\Mb(E_\Mb f,\phi) = \int_\Mb \phi(p)f(p)\dvol_\Mb(p) 
\end{equation}
for $f\in\CoinX{\Mb}$, $\phi\in\Sol(\Mb)$, which implies
\begin{equation}
\sigma_\Mb(E_\Mb f,E_\Mb f') = E_\Mb(f,f')
\end{equation}
and also shows that $\sigma_\Mb$ is weakly nondegenerate.

Now suppose a morphism $\psi:\Mb\to\Nb$ is given, and define the push-forward on 
test functions by $\psi_*f$ 
\[
(\psi_*f)(p)= \begin{cases} f(\psi^{-1}(p)) & p\in\psi(\Mb)\\
0 & \textrm{otherwise,}\end{cases}
\]
for $f\in\CoinX{\Mb}$. As $\Sol(\Mb) = E_\Mb\CoinX{\Mb}$ and
\[
\psi_*\ker E_\Mb =
\psi_* P_\Mb\CoinX{\Mb}=P_\Nb\psi_*\CoinX{\Mb}\subset \ker E_\Nb,
\]
there is a unique linear map $\Sol_\Kbb(\psi):\Sol(\Mb)\to\Sol(\Nb)$, with 
$\Sol(\psi)\circ E_\Mb = E_\Nb\circ\psi_*$ which extends a solution on $\Mb$ to one on $\Nb$. Covariance of the
field equation together with uniqueness of advanced/retarded solutions to the inhomogeneous Klein--Gordon equation
gives the identity $\psi^*E_\Nb\psi_*=E_\Mb$ on $\CoinX{\Mb}$ and it follows that
$\psi^*\Sol(\psi)\phi = \phi$ for all $\phi\in \Sol(\Mb)$, so $\Sol(\psi)$ is injective.

Owing to the calculation
\begin{align*}
\sigma_\Nb(\Sol(\psi)E_\Mb f,\Sol(\psi)E_\Mb f') &= \sigma_\Nb(E_\Nb\psi_*f,E_\Nb\psi_*f') =
E_\Nb(\psi_*f,\psi_*f') = E_\Mb(f,f') \\
&= \sigma_\Mb(E_\Mb f,E_\Mb f')
\end{align*}
and the obvious properties in relation to compositions and identities, $\Sol$ is a functor from $\Mand$ to $\preSympl_\CC$; as $\Sol(\Mb)$ is weakly nondegenerate for every $\Mb$, we
describe $\Sol$ as weakly nondegenerate. Accordingly, there is a unique factorization of $\Sol$
through the forgetful functor from $\Sympl_\CC$ to $\preSympl_\CC$. 

\paragraph{Time-slice property and relative Cauchy evolution}

The following result is a mild extension of a standard argument \cite[Lem. A.3]{Dimock1980}.
\begin{Lem} \label{lem:Pchi}
(i) Let $K$ be a compact subset of $\Mb$, with $K\subset O\in\OO(\Mb)$.
Then there exists a smooth function $\chi$ such
that every $\phi\in\Sol(\Mb)$ with $\supp \phi\subset J_\Mb(K)$ may be written as
$\phi = E_\Mb P_\Mb\chi \phi$ with $P_\Mb\chi\phi\in \CoinX{O}$. (ii) If $O\in\OO(\Mb)$
contains a Cauchy surface of $\Mb$, then there exists a smooth function $\chi$ such
that every $\phi\in\Sol(\Mb)$ may be written as
$\phi = E_\Mb P_\Mb\chi \phi$ with $P_\Mb\chi\phi\in \CoinX{O}$.
\end{Lem}
{\noindent\em Proof:} (i) As $K$ is compact and contained in $O$, $J_\Mb(K)$ has compact intersection with Cauchy surfaces of $O$ (using e.g.,  \cite[Cor. A.5.4]{BarGinouxPfaffle} and causal
convexity of $O$). Choosing Cauchy surfaces $\Sigma^\pm$ of $O$ passing to the future $(+)$ and 
past ($-$) of $K$, the sets $K^\pm=J_\Mb(K)\cap\Sigma^\pm$ are compact and give a cover
\[
J_\Mb(K) = J_\Mb^+(K^+) \cup J_\Mb^-(K^-) \cup K_0
\]
in which the first two sets on the right-hand side are closed and disjoint, while
$K_0 = J_\Mb(K)\cap (J_\Mb^-(K^+)\cap J_\Mb^+(K^-))$ is compact  (see
e.g., \cite[Lem. A.5.7]{BarGinouxPfaffle}) and contained in $O$. 
We may therefore choose $\chi\in C^\infty(\Mb)$ with $\chi=0$ on $J_\Mb^+(K^+)$ and $\chi=1$ on $J_\Mb^-(K^-)$ (see, e.g., \cite[Prop. 5.5.8]{AbrahamMarsdenRatiu}). Following a standard argument \cite{Dimock1980}, $P_\Mb\chi\phi$ is supported in the compact set $K_0\subset O$; on support grounds, it follows that 
\[
\chi\phi=E_\Mb^- P_\Mb\chi\phi,\qquad 
(\chi-1)\phi=E_\Mb^+ P_\Mb\chi\phi
\]
and hence $\phi = E_\Mb P_\Mb\chi\phi\in E_\Mb \CoinX{O}$. 

(ii) If $O\in\OO(\Mb)$ contains a Cauchy surface $\Sigma$ of $\Mb$, then it also contains
Cauchy surfaces $\Sigma^\pm$ passing to the future/past of $\Sigma$. As $J_\Mb^+(\Sigma^+)$ 
and $J_\Mb^-(\Sigma^-)$ are closed and disjoint, we may choose 
$\chi\in C^\infty(\Mb)$ with $\chi=0$ on $J_\Mb^+(\Sigma^+)$ and $\chi=1$ on $J_\Mb^-(\Sigma^-)$; the argument proceeds as before.  $\square$

It follows immediately that $\Sol$ has the time-slice property: if $\psi:\Lb\to\Mb$ is
Cauchy,  then $\Sol(\psi)$ is surjective in addition to being symplectic and injective and therefore has 
a symplectic inverse. Moreover, because $\psi(\Lb)$ contains a Cauchy surface of $\Mb$ we may also
characterize $\Sol(\psi)\phi$ as the unique $P_\Mb$-solution on $\Mb$ whose pull-back to $\Lb$
coincides with $\phi$. This allows us to read off the relative Cauchy evolution.
Given $\hb\in H(\Mb)$ and setting $\Mc^\pm=\Mb\setminus J^\mp_\Mb(\supp\hb)$, 
\[
\rce_\Mb[\hb]\phi=\Sol(\imath^-_\Mb[\hb])\circ\Sol(\jmath^-_\Mb[\hb])^{-1}
\circ \Sol(\jmath^+_\Mb[\hb])\circ \Sol(\imath^+_\Mb[\hb])^{-1} \phi\in\Sol(\Mb)
\]
is the unique $P_\Mb$-solution agreeing on $\Mc^-$ with the unique
$P_{\Mb[\hb]}$-solution that agrees with $\phi$ on $\Mc^+$. Explicit formulae are given in Sec.~4 of BFV (notation differs) 
and Appendix~\ref{appx:diff}.

An important example arises in the massless case. If $\Mb$ has
one or more components with compact Cauchy surfaces, there are nontrivial solutions $\phi$ which are locally constant, i.e., take constant (possibly different) values on each connected component. 
These are solutions to the massless Klein--Gordon equation for {\em any} smooth metric on the 
underlying manifold of $\Mb$ and are therefore fixed points under {\em arbitrary} relative Cauchy evolution. 

As first pointed out by BFV, the functional derivative of the relative Cauchy evolution with respect
to the metric is closely related to the stress energy tensor. In the present setting this can be 
seen as follows.  Let $\Sym(\Mb)$ denote the space of smooth symmetric second rank covariant tensor fields of compact 
support on $\Mb$, and $\Sym(\Mb;O)$ the subspace consisting of those supported in
$O\subset\Mb$. For $\fb\in\Sym(\Mb)$, $s\mapsto \rce_\Mb[s\fb]$ is differentiable at $s=0$
in the weak symplectic topology, i.e., 
there exists a linear map $F_\Mb[\fb]:\Sol(\Mb)\to\Sol(\Mb)$ such that
\begin{equation}\label{eq:wsdiff}
\sigma_\Mb(F_\Mb[\fb]\phi,\phi')= \left.\frac{d}{ds} \sigma_\Mb(\rce_\Mb[s\fb]\phi,\phi')\right|_{s=0}\qquad (\phi\in\Sol(\Mb))
\end{equation}
for any $\fb\in\Sym(\Mb)$. The maps $F_\Mb[\fb]$ are given in BFV:\footnote{
Although BFV give a formula for the derivative, the precise sense in which differentiability is understood was not 
delineated there, nor was differentiability actually proved. This is remedied here; see Appendix~\ref{appx:diff} for a proof of differentiability.
The weak symplectic topology plays a role in general investigations of the canonical commutation relations~\cite{ManVer1968} and appears
the natural choice here.}
\begin{align}
F_\Mb[\fb]\phi &= E_\Mb\left( 
\frac{1}{2}\left(\nabla^a
f{}^b{}_b\right)\nabla_a\phi
-\nabla_a f^{ab} \nabla_b\phi\right) \notag \\
&=
E_\Mb\left( 
\frac{1}{2}\nabla^a 
f{}^b{}_b\nabla_a\phi
-\nabla_a f^{ab} \nabla_b\phi+\frac{1}{2}m^2\phi f{}^b{}_b\right)
\label{eq:FMdef}
\end{align}
where the Klein--Gordon equation was employed in the last step. 
However we note that a more informative form can be given: it turns out that
\begin{equation}\label{eq:magic}
\sigma_\Mb(F_\Mb(\fb) \phi,\overline{\phi})= \int_\Mb f_{ab} T_\Mb^{ab}[\phi] \dvol_\Mb,
\end{equation}
where $\Tb_\Mb[\phi]$ is the classical stress-energy tensor on
$\Mb$ for the solution $\phi$:
\begin{equation}
T_\Mb^{ab}[\phi]=(\nabla^{(a}\overline{\phi})(\nabla^{b)}\phi)
-\frac{1}{2}g^{ab}g^{cd}(\nabla_c\overline{\phi})(\nabla_d \phi)
+\frac{1}{2}m^2|\phi|^2 g^{ab}.
\end{equation}
To see this, we use \eqref{eq:sigma_and_E_1} to note that
\[
\sigma_\Mb(F_\Mb(\fb)\phi,\overline{\phi}) = \int_\Mb \overline{\phi}
\left( 
\frac{1}{2}\nabla^a 
f{}^b{}_b\nabla_a\phi -\nabla_a f^{ab} \nabla_b\phi+\frac{1}{2}m^2\phi f{}^b{}_b\right) \dvol_\Mb
\]
and then integrate by parts in the first two terms, using the fact that $\fb$ is compactly
supported and symmetric to discard boundary terms and thereby obtain the required result. 

Let us also recall that if $u$ is a timelike unit vector then
\[
T_\Mb^{ab}[\phi]u_a u_b = \frac{1}{2} h^{ab} \nabla_a\overline{\phi}\nabla_b\phi + \frac{1}{2}
m^2|\phi|^2,
\]
where $h^{ab} = 2u^a u^b- g^{ab}$ is positive definite. Accordingly, vanishing of $\Tb_\Mb[\phi]$ at a point $p$ implies that $\nabla\phi$ vanishes there; for $m>0$ we may also conclude that $\phi$ also vanishes at $p$.

\paragraph{Dynamical locality}

An immediate consequence of the definition is that the kinematic subspaces are given, for nonempty $O\in\OO(\Mb)$, by 
\[
\Sol^\kin(\Mb;O) = \Ran(\Sol(\iota_{\Mb;O})) = E_\Mb \CoinX{O}.
\]

As a slight digression, which will be useful later, we note that if $O, O'\in\OO(\Mb)$ are 
nonempty and causally disjoint, then
\begin{equation}
\sigma_\Mb(\Sol^\kin(\Mb;O), \Sol^\kin(\Mb;O')) = \{0\}
\end{equation}
because $E_\Mb$ vanishes on pairs of test functions with causally disjoint supports. 
By analogy with the situation in algebraic quantum field theory, we call this {\em Einstein causality}. 
It follows that we also have the analogue of the {\em extended locality} property of QFT~\cite{Schoch1968, Landau1969}, 
\[
\Sol^\kin(\Mb;O_1)\cap\Sol^\kin(\Mb;O_2) = \{0\}.
\]
{}For suppose $\phi\neq 0$ is an element of the intersection, then it can be written
as $\phi=\Sol(\iota_{\Mb;O})\hat{\phi}$ for some $\hat{\phi}\in\Sol(\Mb|_O)$. 
By weak nondegeneracy, there exists $\hat{\phi}'\in\Sol(\Mb|_O)$ with
$\sigma_{\Mb|_O}(\hat{\phi},\hat{\phi}')\neq 0$. But then also $\sigma_\Mb(\phi,
\Sol(\iota_{\Mb;O})\hat{\phi}')\neq 0$, which is a contradiction because $\phi$ may
also be regarded as an element of $\Sol^\kin(\Mb;O')$. 

We now proceed to compute the dynamical subspaces. 

\begin{Prop} \label{prop:Lbullet}
Let $K$ be any compact subset of $\Mb\in\Loc$. Then
\[
\Sol^\bullet(\Mb;K) = \{\phi\in\Sol(\Mb):\supp\Tb_\Mb[\phi]\subset J_\Mb(K)\}.
\]
\end{Prop}
{\noindent\em Proof:} Suppose  that $\phi\in\Sol^\bullet(\Mb;K)$. 
Given any $\fb\in\Sym(\Mb;K^\perp)$, there is an interval containing $s=0$ for which $s\fb\in H(\Mb;K^\perp)$; as $\rce_\Mb[s\fb]\phi=\phi$ for all such $s$ we may
differentiate to obtain $F_\Mb[\fb]\phi=0$ for all such $\fb$. It follows immediately from~\eqref{eq:magic} that  $\Tb_\Mb[\phi]$ is supported in $J_\Mb(K)$. 

Conversely, suppose $\phi\in \Tb_\Mb[\phi]$ vanishes in $K^\perp$.
Then $\nabla\phi$ vanishes in $K^\perp$ and so $\phi$ is constant in each connected component of $K^\perp$. Accordingly, $\phi$ is also a Klein--Gordon solution with respect to any perturbed metric 
induced by $\hb\in H(\Mb;K^\perp)$, which shows that $\rce_\Mb[\hb]\phi=\phi$ for all such $\hb$, i.e., 
$\phi\in\Sol^\bullet(\Mb;K)$. $\square$

Note that $\Sol^\bullet(\Mb;K)$ includes solutions whose support extends to the boundary of $J_\Mb(K)$. 
By contrast, if $O$ is a nonempty open relatively compact globally hyperbolic subset of $\Mb$, 
solutions in $\Sol^\kin(\Mb;O)$ have (closed) support contained in $J_\Mb(O)$ which is
a proper subset of $J_\Mb(\cl(O))$. Thus we see that $\Sol^\kin(\Mb;O)$ is a proper
subset of $\Sol^\bullet(\Mb;\cl(O))$ in this case. In general, we also have

\begin{Lem} \label{lem:Lkindyn}
$\Sol^\kin(\Mb;O)\subset \Sol^\dyn(\Mb;O)$ for all nonempty $O\in\OO(\Mb)$. 
\end{Lem}
{\noindent\em Proof:} Suppose $\phi=\Sol^\kin(\Mb;O)$, so $\phi=E_\Mb f$ for
some $f\in \CoinX{O}$. We may decompose $f$ as a finite 
sum $f=\sum f_i$ in which $\supp f_i\in\KK(\Mb;O)$. (Take an open cover of $\supp f$ by diamonds and pass to a finite subcover and then a subordinate partition of unity.) 
Each $E_\Mb f_i$ has support in $J_\Mb(\supp f_i)$ and hence belongs to $\Sol^\bullet(\Mb;\supp f_i)$,
which shows that $\phi =\sum_i E_\Mb f_i\in \Sol^\dyn(\Mb;O)$. $\square$

At this stage, the mass parameter $m$ becomes important. If $m>0$ then, 
using Prop.~\ref{prop:Lbullet} and Lem.~\ref{lem:Pchi}(i),
we have
\[
\Sol^\bullet(\Mb;K)=\{\phi\in\Sol(\Mb): \supp \phi\subset J_\Mb(K)\} \subset 
E_\Mb \CoinX{O} = \Sol^\kin(\Mb;O)
\]
for every $K\in \KK(\Mb;O)$ and $O\in \OO(\Mb)$. Taking a union over all such $K$, 
we obtain the reverse inclusion to Lem.~\ref{lem:Lkindyn}. As dynamical
locality of $\Sol=\Sol_\CC$ implies that of $\Sol_\RR$, we have proved: 
\begin{Thm}
The classical Klein--Gordon theory $\Sol_\Kbb$ is dynamically local in $\LCT$ (and hence its restriction to $\Man$ is dynamically local in $\LCTo$) for all $m>0$. 
\end{Thm}

Now consider the case $m=0$. Any function on $\Mb$ that is locally constant (i.e., constant on 
each connected component of $\Mb$) satisfies the field equation; we denote by $\Sol_{l.c.}(\Mb)$ the space of locally constant solutions with compact support on Cauchy surfaces, which has dimension
equal to the number of compact connected components of $\Mb$. (In the case where $\Mb$ has purely noncompact Cauchy surfaces $\Sol_{l.c.}(\Mb)$ is trivial.) Now from  Prop.~\ref{prop:Lbullet}, we know that 
\[
\Sol^\bullet(\Mb;K)= \{\phi\in \Sol(\Mb):\supp \nabla\phi\subset J_\Mb(K)\}
\]
for any compact set $K$, so any $\phi\in \Sol^\bullet(\Mb;K)$ is constant on each connected
component of $K^\perp$. This allows us to prove:
\begin{Lem} \label{lem:Kperp}
If $K\subset \Mb$ is compact, and $K^\perp$ has connected intersection with each connected component of $\Mb$, then 
\[
\Sol^\bullet(\Mb;K)= \{\phi\in \Sol(\Mb):\supp\phi\subset J_\Mb(K)\} + \Sol_{l.c.}(\Mb).
\]
\end{Lem}
{\noindent\em Proof:} The remarks above and the hypothesis on $K$ permit us to write
any $\phi\in \Sol^\bullet(\Mb;K)$ as a sum $\phi=\phi_{l.c.} + \phi_0$, where
$\phi_{l.c.}\in \Sol_{l.c.}(\Mb)$ is locally constant and $\phi_0$ is supported in $J_\Mb(K)$. 
This gives the inclusion of the left-hand side in the right; for the reverse inclusion, 
we use the fact that every element
of $\Sol_{l.c.}(\Mb)$ is invariant under arbitrary classical relative Cauchy evolution. $\square$

If $O\in\OO(\Mb)$ is nonempty, $\Sol^\dyn(\Mb;O)$ is the span of the subspaces $\Sol^\bullet(\Mb;K)$
for $K\in\KK_b(\Mb;O)$, defined before Def.~\ref{def:dynloc}. In spacetime dimension $n>2$ each such $K$ meets the hypotheses of Lem.~\ref{lem:Kperp} 
and we easily see that 
\[
\Sol^\dyn(\Mb;O) = \Sol^\kin(\Mb;O) + \Sol_{l.c.}(\Mb).
\]
Spacetime dimension $n=2$ is complicated by the fact that every nonempty $K\in\KK_b(\Mb;O)$ has
disconnected intersection with at least one of the connected components of $\Mb$. 
This is worthy of more discussion, but
for the present, we simply observe that $\Sol^\dyn(\Mb;O)$ contains $\Sol^\kin(\Mb;O) + \Sol_{l.c.}(\Mb)$. 

Summarising, we have shown:
\begin{Thm} 
The classical Klein--Gordon theory $\Sol_\Kbb$ is not dynamically local for $m=0$.
\end{Thm}

The classical massless Klein--Gordon field is singled out by not obeying the dynamical locality property. 
As we will see below, this also propagates to the quantum field theory. Although the discrepancy between the kinematic and dynamical subspaces is slight and under full control, we adopt the viewpoint that the failure of dynamical locality should be taken seriously as an 
indication of a defect in the usual treatment of the massless minimally coupled model. The root cause is easily seen: namely, the rigid gauge symmetry $\phi\mapsto \phi+\textrm{const}$ in the Lagrangian. 
In the next section, we show how the massless theory can be formulated in a dynamically local way, by treating it as a (rather simple) gauge theory. 

To conclude this section, we summarise a number of features of the Klein--Gordon theory with
arbitrary mass $m\ge 0$ from the above discussion. For $\Sol=\Sol_\CC$ or $\Sol_\RR$ we have:
\begin{itemize}\addtolength{\itemsep}{-0.5\baselineskip}
\item[($\Sol1$)] $\Sol$ 
has a smooth stress-energy tensor, i.e., the relative Cauchy evolution is differentiable in the weak symplectic topology 
as in \eqref{eq:wsdiff}, and the resulting maps $F_\Mb[\fb]$ obey 
\begin{equation}
\sigma_\Mb(F_\Mb[\fb]\phi, C_\Mb\phi) = \int f_{ab} T_\Mb^{ab}[\phi]\dvol \qquad (\fb\in\Sym(\Mb;O)),
\end{equation}
where $\Tb_\Mb[\phi]\in C^\infty(T^{2}_0(\Mb))$ is a smooth conserved symmetric tensor field
for each $\phi\in\Sol(\Mb)$ (in the case $\Kbb=\RR$, take $C_\Mb$ to be the identity). 
\item[($\Sol2$)] For each $O\in\OO(\Mb)$ containing $\supp \fb$, 
we have $\im F_\Mb[\fb]\subset \Sol^\kin(\Mb;O)$.
\item[($\Sol3$)] $\Sol$ obeys  extended locality. 
\item[($\Sol4$)] The stress-energy tensor is sufficient to define the dynamical subspaces, i.e.,
\begin{equation}
\Sol^\bullet(\Mb;K) = \bigcap_{\fb\in\Sym(\Mb;K^\perp)} \ker F_\Mb[\fb].
\end{equation}
\end{itemize}
One might expect these properties to hold for wide range of theories of interest. Later, it
will be useful to have the following technical result, in which dynamical locality is not assumed.
\begin{Prop} \label{prop:inv_to_bullet}
Suppose $\Sol:\Mand\to\preSympl_\Kbb$  is weakly nondegenerate and obeys properties ($\Sol1$--$\Sol4$). Let $O$ be a nonempty open subset of $\Mb\in\Mand$. If a finite-dimensional subspace $Y\subset\Sol(\Mb)$ is invariant under $F_\Mb[\fb]$ for every $\fb\in\Sym(\Mb;O)$  then $Y\subset 
\bigcap_{\fb\in \Sym(\Mb;O)}\ker F_\Mb[\fb]$. (In particular, every $\phi\in Y$ has vanishing 
stress-energy tensor in $O$.) In the case $O=K^\perp$ for compact $K$, this implies that $Y\subset 
\Sol^\bullet(\Mb;K)$.
\end{Prop}
{\noindent\em Remark:} In particular, we any finite dimensional subspace $Y$ that is invariant under $\rce_\Mb[\hb]$ for all $\hb\in H(\Mb;K^\perp)$ obeys $Y\subset \Sol^\bullet(\Mb;K)$. The result also holds with $\Mand$ replaced by $\Man$ throughout. \\
{\em Proof.}
Let $p\in O$ be arbitrary. In any neighbourhood $\tilde{O}$ of $p$ with $\tilde{O}\subset O$ choose causally
disjoint open subsets $O_1,\ldots,O_{1+\dim Y}$ and consider the
subspaces
\begin{equation}
Y_i = \bigvee_{\fb\in\Sym(\Mb;O_i)} \im F_\Mb[\fb]|_Y
\end{equation}
of $Y$ (where $\bigvee$ denotes the span of subspaces). Using  ($\Sol2$) and ($\Sol3)$, it is
clear that the $Y_i$ constitute $1+\dim Y$ 
subspaces of $Y$ with trivial pairwise intersections,so
at least one of the $Y_i$, say $Y_1$, must be trivial. For each $\phi\in Y$ we must
therefore have $F_\Mb[\fb]\phi=0$ for every $\fb\in \Sym(\Mb;O_1)$. 

The existence of a smooth stress-energy tensor entails, by a polarisation argument, that 
for each $\phi'\in \Sol(\Mb)$ there is a smooth tensor field $\Tb_\Mb[\phi,\phi']$ with
 $\sigma_\Mb(F_\Mb[\fb]\phi,C_\Mb\phi')= \int_\Mb T^{ab}_\Mb[\phi,\phi'] f_{ab}\dvol$, where $C_\Mb$ is the conjugation in $\Sol(\Mb)$. 
The above remarks show that $\Tb_\Mb[\phi,\phi']$  vanishes
identically in $O_1\subset \tilde{O}$. As $\tilde{O}$ was arbitrary, it follows
by continuity that $\Tb_\Mb[\phi,\phi']$ vanishes
at $p$, which was an arbitrary point of $O$. 
Consequently, $\sigma_\Mb(F_\Mb[\fb]\phi,C_\Mb\phi')=0$  for arbitrary $\phi'\in\Sol(\Mb)$, $\fb\in\Sol(\Mb;O)$. As $\sigma_\Mb$ is weakly nondegenerate, we conclude that $F_\Mb[\fb]\phi=0$ for all $\phi\in Y$ and $\fb\in\Sym(\Mb;O)$.  The last statement follows from ($\Sol4$). $\square$

\section{The massless current}\label{sect:currents}

Our eventual aim is to quantize the massless minimally coupled model as a gauge theory, following the general
lines of treatments of the electromagnetic field \cite{Dimock92, Few&Pfen03} or its analogues \cite{Pfenning09} 
(although we will {\em not} make the cohomological restrictions imposed in these references). 
In this section, we describe the underlying classical field theory, aiming for a dynamically local
theory valued in $\preSympl_\Kbb$. As before, we will focus on the complex case, dropping $\CC$
from the notation, with $\Sol$ denoting the massless Klein--Gordon field
as formulated above (in $\preSympl_\CC$). 

It is convenient to employ differential forms: as usual, $d_\Mb$ denotes the exterior derivative, while 
$\delta_\Mb$ is the codifferential, defined with the conventions of \cite{AbrahamMarsdenRatiu} (also used in \cite{Few&Pfen03}) in which $\Box_\Mb = -(\delta_\Mb d_\Mb + d_\Mb\delta_\Mb)$,
[i.e., minus the Laplace--de Rham operator], which agrees with the action of 
$g^{ab}\nabla_a\nabla_b$ up to lower order terms that vanish on $0$-forms. 
Thus $\Box_\Mb$ has metric principal symbol on $p$-forms of any degree and has unique
advanced ($-$) and retarded ($+$) fundamental solutions $E_\Mb^\pm:\Omega_0^p(\Mb)\to\Omega^p(\Mb)$ 
extending the usual notation for $0$-forms. A key property is
that the exterior derivative and coderivative commute with 
$\Box_\Mb$ and hence $E^\pm_\Mb$ (or more precisely intertwine their actions on forms of adjacent rank).

{}For any $\Mb\in\Mand$, let $\Sol_0(\Mb)$ be the space of $\phi\in C^\infty(\Mb)$ such that
$\Box_\Mb\phi=0$ and obeying the following conditions: (A) there is at least one locally constant function $c$ so that the support of $\phi-c$ has compact intersection with 
all Cauchy surfaces;\footnote{The function $c$ is unique if and only if $\Mb$ has purely noncompact Cauchy surfaces.} 
(B) the constraint
\begin{equation}\label{eq:charge_zero}
\int_\Sigma \nabla_n \phi d\Sigma = 0
\end{equation}
holds on each smooth spacelike Cauchy surface $\Sigma$ of each component of $\Mb$, where $n^a$ is the unit future-pointing normal field to $\Sigma$. It is enough 
to verify conditions (A) and (B) for any particular choices of Cauchy surface to deduce that they hold in general:
for (B) this follows using the field equation and divergence theorem, in conjunction with the support 
properties of (A). Conditions (A) and (B) together entail
that the usual formula for the symplectic product $\sigma_\Mb$ gives a well-defined convergent
integral on solutions in $\Sol_0(\Mb)$, although it is now degenerate as the locally constant solutions
have vanishing symplectic product with all other solutions -- this is exactly the content of Eq.~\eqref{eq:charge_zero},
which can also be interpreted as the vanishing of the Noether charge associated with the $\phi\mapsto\phi+c$ invariance in each component of $\Mb$. 
Next, we define the linear equivalence relation
$\phi\sim\phi'$ on $\Sol_0(\Mb)$ to mean that $\phi-\phi'$ is locally constant, and then take the
quotient $\Jf(\Mb)=\Sol_0(\Mb)/\sim$ as the classical phase space of the theory. 
Condition (B) guarantees that the symplectic form descends to an antisymmetric bilinear form $\sigma_{0\,\Mb}$ on 
the equivalence classes in $\Jf(\Mb)$ and is readily seen to be nondegenerate: if
$\sigma_{0\,\Mb}([\phi],[\phi']) = 0$ for all $[\phi']\in \Jf(\Mb)$, then 
$\sigma_{\Mb}(\phi,\phi') = 0$ for all $\phi'\in \Sol(\Mb)$, where $\phi$ is a representative
of $[\phi]$ in $\Sol(\Mb)$; it follows that $\phi$ and hence $[\phi]$ vanish, because $\sigma_\Mb$
is weakly nondegenerate. Thus $\Jf(\Mb)$, equipped with $\sigma_{0\,\Mb}$ and complex
conjugation defines a weakly nondegenerate object of $\preSympl_\CC$. 

The covariance of this theory is easily established using the following result. 
\begin{Lem} \label{lem:current_covariance}
Suppose $\psi:\Mb\to\Nb$ in $\Mand$. Then (a) $\Sol(\psi)(\Sol_{l.c.}(\Mb))\subset \Sol_{l.c.}(\Nb)$; (b) if $\phi\in\Sol(\Mb)$ obeys condition (B) in $\Mb$ (whereupon $[\phi]\in\Jf(\Mb)$) then $\Sol(\psi)\phi$ also obeys condition (B) in $\Nb$ and hence $[\Sol(\psi)\phi]\in\Jf(\Nb)$. 
\end{Lem}
{\noindent\em Proof}  (a) Take $\chi\in C^\infty(\Mb)$ so that
$d\chi=0$ outside a neighbourhood of a Cauchy surface in $\Mb$ and so that
$\eta=E_\Mb\Box_\Mb\chi\eta$ for any $\eta\in\Sol(\Mb)$. Then $\Sol(\psi)\eta = E_\Nb\psi_*\Box_\Mb\chi\eta$ and using well-known intertwining relations,
\[
d\Sol(\psi)\eta=E^{(1)}_\Nb d \psi_*\Box_\Mb \chi\eta
=E^{(1)}_\Nb  \psi_*\Box_\Mb^{(1)} d \chi\eta,
\]
where
$d$ is the exterior derivative and $E^{(1)}$, $\Box^{(1)}$ are the $1$-form analogues of $E$, $\Box$. If $\eta\in\Sol_{l.c.}(\Mb)$ then $d\chi\eta$ has
compact support\footnote{Recall that $\eta$ can only be nonzero on components of $\Mb$ with
compact Cauchy surfaces.} and we get $\psi_*\Box_\Mb^{(1)} d \chi\eta=
\Box_\Nb^{(1)} \psi_* d \chi\eta$, whereupon it is clear that $d\Sol(\psi)\eta=0$
so $\Sol(\psi)\eta\in \Sol_{l.c.}(\Nb)$. 

(b) We observe that $\phi = E_\Mb f$ ($f\in\CoinX{\Mb}$) obeys condition (B) if and only
if for each component $\Bb$ of $\Mb$, with Cauchy surface $\Sigma_\Bb$,
\[
0=\int_{\Sigma_\Bb} \nabla_n \phi d\Sigma = \int_\Bb f\,\dvol_\Bb
\]
using~\cite[Lem.~A.1]{Dimock1980} applied to the smooth constant solution $1$. 
Thus if $\phi=E_\Mb f$ obeys condition (B) then $\Sol(\psi)\phi=E_\Nb \psi_* f$ with $\int_\Cb \psi_*f\dvol =0$
in each component $\Cb$ of $\Nb$, so $\Sol(\psi)\phi$ obeys condition (B) in $\Nb$. $\square$

Given this result, for $\psi:\Mb\to\Nb$ it is possible to define $\Jf(\psi):\Jf(\Mb)\to\Jf(\Nb)$ by $\Jf(\psi)[\phi] = 
[\Sol(\psi)\phi]$, where $\phi$ is a representative lying in $\Sol(\Mb)$. (Note that
$\Sol(\psi)\phi$ is then a representative of $\Jf(\psi)[\phi]$ in $\Sol(\Nb)$.) 
This is well-defined because if $[\phi]=[\phi']$ with both
$\phi,\phi'\in \Sol(\Mb)$, then $\phi-\phi'\in\Sol_{l.c.}(\Mb)$ and hence
$\Sol(\psi)\phi\sim\Sol(\psi)\phi'$ by Lemma~\ref{lem:current_covariance}(a).
Moreover, as 
\[
\sigma_{0\,\Nb}(\Jf(\psi)[\phi],\Jf(\psi)[\phi']) = \sigma_{\Nb}(\Sol(\psi)\phi,\Sol(\psi)\phi') 
=\sigma_{\Mb}(\phi,\phi') = \sigma_{0\,\Mb}([\phi],[\phi']),
\]
it is clear that $\Jf(\psi):\Jf(\Mb)\to\Jf(\Nb)$ in $\Sympl$, so $\Jf(\psi)$ is
necessarily injective. The 
functorial property of $\Sol$ induces the corresponding property for $\Jf$. 
Hence $\Jf$ is indeed a functor from $\Loc$ to $\Sympl$ and is easily shown
to have the timeslice property: if $\psi$ is Cauchy, it is clear that $[\phi]\mapsto [\Sol(\psi)^{-1}\phi]$
is inverse to $\Jf(\psi)$ -- one need only check that $\Sol(\psi)$ maps
$\Sol_{l.c.}(\Mb)$ surjectively onto $\Sol_{l.c.}(\Nb)$ for Cauchy $\psi$. 

{}From the definition of $\Jf(\psi)$ it follows immediately that the kinematic
subspaces are given as
\[
\Jf^\kin(\Mb;O) = [\Sol^\kin(\Mb;O)]
\]
for nonempty $O\in\OO(\Mb)$. Furthermore, the relative Cauchy evolution is
easily seen to be 
\[
\rce_{0\,\Mb}[\hb] [\phi] = [\rce_{\Mb}[\hb]\phi]
\]
for $\hb\in H(\Mb)$, where $\phi$ is a representative in $\Sol(\Mb)$ of
$[\phi]$, and we use $\rce_0$ and $\rce$ in place of $\rce^{(\Jf)}$ and $\rce^{(\Sol)}$ to unburden the notation.

\begin{Prop} \label{prop:Sol0_bullet_dyn_kin}
For any compact $K\subset\Mb$, and any $O\in\OO(\Mb)$ we have
\[
\Jf^\bullet(\Mb;K) = [\Sol^\bullet(\Mb;K)], \qquad\text{and}\qquad
\Jf^\dyn(\Mb;O) = [\Sol^\dyn(\Mb;O)].
\]
If, additionally, $O$ is nonempty, we also have $\Jf^\kin(\Mb;O)\subset \Jf^\dyn(\Mb;O)$.
\end{Prop}
{\noindent\em Proof:} 
We note that $[\phi]\in\Jf^\bullet(\Mb;K)$
(with $\phi$ a representative in $\Sol(\Mb)$) if and only if $\rce_\Mb[\hb]\phi\sim
\phi$ for all $\hb\in H(\Mb;K^\perp)$. But as both $\phi$ and $\rce_\Mb[\hb]\phi$
have compact support on Cauchy surfaces, their difference must therefore be
an element of $\Sol_{l.c.}(\Mb)$. Accordingly 
the finite dimensional subspace $\CC\phi + \Sol_{l.c.}(\Mb)$ is invariant
under all such $\rce_\Mb[\hb]$, and by Prop.~\ref{prop:inv_to_bullet}
is therefore contained in $\Sol^\bullet(\Mb;K)$.
In particular, $\phi\in\Sol^\bullet(\Mb;K)$. As the reverse inclusion is trivial, the first
equality is proved;  the second follows immediately on taking the linear spans. 
Finally, we use the above results to compute
\[
\Jf^\kin(\Mb;O) =  [\Sol^\kin(\Mb;O)] \subset [\Sol^\dyn(\Mb;O)] = \Jf^\dyn(\Mb;O),
\]
where the inclusion follows from Lem.~\ref{lem:Lkindyn}.
$\square$

The spacetime dimension now enters in an essential way. In dimensions $n>2$,
$\Sol^\dyn(\Mb;O)$ differs from $\Sol^\kin(\Mb;O)$ (for nonempty $O\in\OO(\Mb)$) only by locally constant solutions, which are annihilated by the quotient, thus giving dynamical locality:
\[
\Jf^\dyn(\Mb;O) = [\Sol^\dyn(\Mb;O)] = [\Sol^\kin(\Mb;O)+\Sol_{l.c.}(\Mb)] 
= [\Sol^\kin(\Mb;O)] = \Jf^\kin(\Mb;O).
\]
In $n=2$ dimensions, however, there is an added complication. First, consider the case in which $O\in\OO_0(\Mb)$
for $\Mb\in\Man$. If $K\subset O$ is the closure of the base of a multi-diamond, then there is a smooth spacelike Cauchy surface $\Sigma$ for $O$ that contains $K$. As a connected one-dimensional paracompact manifold, $\Sigma$ is homeomorphic to either $\RR$ or $S^1$; in either case, as $K$ must have nonempty causal complement, it is evident that we may find a connected, contractible compact set $\tilde{K}$ so that $K\subset \tilde{K}\subset \Sigma$. Then it is clear that
\[
\Jf^\bullet(\Mb;K)\subset \Jf^\bullet(\Mb;\tilde{K}) \subset \Jf^\kin(\Mb;O)
\]
and as $\Jf^\dyn(\Mb;O)$ is generated over such $K$, we have the inclusion
$\Jf^\dyn(\Mb;K)\subset  \Jf^\kin(\Mb;O)$. Together with the last 
statement of Prop.~\ref{prop:Sol0_bullet_dyn_kin}, this establishes dynamical locality for $\Jf$ in $\LCTo$, i.e., when regarded as a theory on  the category of connected spacetimes, $\Man$. 

On the other hand, we can also see that dynamical locality fails when disconnected spacetimes are permitted, i.e., in $\LCT$.  Let $\Mb_0$ be two-dimensional Minkowski space, with standard
$(t,x)$ coordinates and metric $dt^2-dx^2$, and let $O$ be the Cauchy development of the set
$B=\{(0,x): a<|x|<2a\}$ for some $a>0$. The causal complement $B^\perp$ consists of
the Cauchy development of $(-\infty,-2a)\cup (-a,a) \cup (2a,\infty)$, which has three connected components. Now let $f\in\CoinX{\RR}$ be such that
$f\equiv 1$ on $(-a,a)$ and $f\equiv 0$ outside $(-2a,2a)$. Then 
$\phi(t,x) =\frac{1}{2}\left( f(x-t) + f(x+t)\right)$ is a solution taking the value $1$ inside the Cauchy development of $(-a,a)$ and vanishing in the
other two components of $B^\perp$. This solution is invariant under relative Cauchy evolutions
induced by metric perturbations supported in $B^\perp$ but is not locally constant; accordingly, 
$[\phi]$ is a nonzero element of $\Jf^\bullet(\Mb_0;B)\subset \Jf^\dyn(\Mb_0;O)$. 
However, $[\phi]$ is not an element of $\Jf^\kin(\Mb_0;O)$, because the pull-back
$\iota_{\Mb;O}^*\phi$ of $\phi$ to $O$ cannot be reduced to a solution of compact support on
Cauchy surfaces (of $O$) by adding a locally constant function. Thus $\Jf^\kin(\Mb_0;O)$ is a 
proper subspace of $\Jf^\dyn(\Mb_0;O)$, and dynamical locality fails. 

Summarising: 
\begin{Thm} The theory $\Jf$ is dynamically local in $\LCTo$ for all dimensions $n\ge 2$. 
It is dynamically local in $\LCT$ for all dimensions $n\ge 3$, but not in dimension $n=2$. 
\end{Thm}
Moreover, the above discussion shows that $\Jf$ obeys conditions ($\Sol1$--$\Sol4$) in
all dimensions $n\ge 2$, and whether formulated in $\LCT$ or $\LCTo$. For
($\Sol1$) and ($\Sol2$) hold because the relative Cauchy evolution of $\Jf$ is inherited from that of $\Sol$, and the classical stress-energy tensor of the massless scalar field is independent of the choice
of representative in equivalence classes modulo locally constant functions; likewise, ($\Sol3$) 
follows from the extended locality of $\Sol$, while ($\Sol4$) holds by the first equation in Prop.~\ref{prop:Sol0_bullet_dyn_kin}.

The failure of dynamical locality for $\Jf$ in $\LCT$ for dimension $n=2$ suggests the need for further work. The cause of the defect is evidently connected to the presence of topological charge; we conjecture that it can be addressed by admitting topological charge as an additional background feature of spacetimes. 
At any rate, it seems clear that when dynamical locality fails, 
it does so for interesting reasons. 

Finally, we give a technical result that will be used later on.
\begin{Lem} \label{lem:current}
The map $\Omega_0^1(\Mb)\owns\omega\mapsto [E_\Mb\delta_\Mb\omega]$ is a linear surjection onto 
$\Jf(\Mb)$ with kernel $\ker\delta_\Mb\cap\Omega_0^1(\Mb) + (\Omega_0^1(\Mb)\cap d_\Mb C^\infty(\Mb))$.
\end{Lem}
{\noindent\em Proof:} First, suppose $\omega\in\Omega_0^1(\Mb)$ and set $\phi=E_\Mb\delta_\Mb\omega$. Let $\Cb$ be any connected component of $\Mb$,
let $\Sigma_\Cb$ be a Cauchy surface for $\Cb$ with unit future-pointing normal $\nb$, and $1_\Cb$ the locally constant function on $\Mb$ that takes the value $1$ on $\Cb$ and $0$ otherwise.
Then 
\[
 \int_{\Sigma_\Cb} \nabla_\nb \phi \,d\Sigma_{\Cb}   = \sigma_\Mb(E_\Mb \delta_\Mb\omega,1_\Cb) =\int_\Mb 1_\Cb \delta_\Mb\omega \dvol_\Mb = 
\int_\Cb  \delta_\Cb\omega|_\Cb \dvol_\Cb =0,
\]
by~\cite[Lem.~A.1]{Dimock1980} (because $1_\Cb$ is a smooth solution while
$\delta_\Mb\omega\in\CoinX{\Mb}$) and as $\Cb$ was arbitrary, we have
$[E_\Mb\delta_\Mb\omega]\in\Jf(\Mb)$; the map is evidently linear. To check surjectivity, 
given any element of $\Jf(\Mb)$ choose a representative $\phi$ with compact support on Cauchy surfaces and
write $\phi= E_\Mb f$ for some $f\in\CoinX{\Mb}$. Then, with $\Cb$, $\Sigma_\Cb$, $1_\Cb$ as before,
\[
\int_\Cb f|_\Cb \dvol_\Cb = \sigma_\Cb(E_\Cb f|_\Cb,1) = \int_{\Sigma_\Cb}\nabla_\nb \phi \,d\Sigma_\Cb = 0,
\]
{}from which we may deduce that $f|_\Cb\in\delta_\Cb\Omega_0^1(\Cb)$ for each $\Cb$
and hence $f\in \delta_\Mb\Omega_0^1(\Mb)$ by a
standard result on the compact cohomology group of highest degree for connected manifolds,
(see,  e.g., Theorem~7.5.19(i) in~\cite{AbrahamMarsdenRatiu}). Accordingly, surjectivity holds.

Next, suppose $\omega\in\Omega_0^1(\Mb)$ and  $[E_\Mb\delta_\Mb\omega]=0$. Then
$0=d_\Mb E_\Mb \delta_\Mb\omega= E_\Mb d_\Mb\delta_\Mb\omega$ and hence
\begin{equation}\label{eq:omega_and_beta}
d_\Mb\delta_\Mb\omega = \Box_\Mb \beta
\end{equation} 
for some $\beta\in\Omega_0^1(\Mb)$. In particular, $\beta$ vanishes to the past, so we may solve 
for $\beta$ by applying the retarded fundamental solution for $1$-forms, 
\[
\beta = E^{+}_\Mb d_\Mb\delta_\Mb\omega =   d_\Mb E^{+}_\Mb \delta_\Mb\omega 
\in \Omega_0^1(\Mb)\cap d_\Mb C^\infty(\Mb)
\]
(equally, we could have used the advanced fundamental solution). 
On the other hand, Eq.~\eqref{eq:omega_and_beta} also implies 
\[
\Box_\Mb\delta_\Mb\omega = -\delta_\Mb d_\Mb\delta_\Mb \omega = -\delta_\Mb\Box_\Mb\beta 
= -\Box_\Mb\delta_\Mb\beta
\]
and hence $\delta_\Mb\omega=-\delta_\Mb\beta$ (as both have compact support). Thus
$\omega=-\beta+\kappa$ for some $\kappa\in \ker\delta_\Mb$ and $\beta\in  \Omega_0^1(\Mb)\cap d_\Mb C^\infty(\Mb)$. 
Finally, suppose that $\omega$ takes this form, with $\beta=-d_\Mb\chi\in  \Omega_0^1(\Mb)$, $\chi\in C^\infty(\Mb)$. 
Then 
\[
d_\Mb E_\Mb \delta_\Mb\omega = E_\Mb d_\Mb\delta_\Mb d_\Mb\chi= 
-E_\Mb\Box_\Mb d_\Mb\chi = 0
\]
(recall that $d_\Mb\chi$ has compact support) so $[E_\Mb\delta_\Mb\omega]=[0]$.  $\square$

\section{Quantized theory: smeared fields}\label{sect:infinitesimal}

In this section, we describe how weakly nondegenerate dynamically local classical theories, valued in $\preSympl_\CC$
and obeying the additional conditions ($\Sol1$--$\Sol4$), can be quantized 
as the `infinitesimal Weyl algebra' (cf.~\cite{BaezSegalZhou}) 
to obtain a dynamically local quantum field theory. For the Klein--Gordon case, this is the usual $*$-algebra of smeared fields. We begin by describing this quantization method in a form that will be convenient for our purposes. We will also
make contact with other standard presentations of the theory.

\subsection{Quantization functor}
 
The infinitesimal Weyl algebra quantization of a presymplectic space $(V,\sigma,C)\in \preSympl_\CC$ is given by
the unital $*$-algebra $\Qf(V,\sigma,C)$, whose underlying complex vector space is the symmetric tensor vector space over $V$,
\begin{equation}\label{eq:Q1}
\Qf(V,\sigma,C) = \Gamma_\odot(V) \stackrel{\text{def}}{=} \bigoplus_{n\in\NN_0} V^{\odot n} ,
\end{equation}
equipped with a product such that
\begin{equation}\label{eq:Q2}
u^{\odot m} \cdot v^{\odot n} = \sum_{r=0}^{\min\{m,n\}}
\left(\frac{i\sigma(u,v)}{2}\right)^r \frac{m!n!}{r!(m-r)!(n-r)!}
S\left( u^{\otimes (m-r)}\otimes v^{\otimes (n-r)}\right),
\end{equation}
where $S$ denotes symmetrisation, and a $*$-operation defined by $(u^{\odot n})^* = (Cu)^{\odot n}$; both
operations being extended by (anti-)linearity to general elements of $\Gamma_\odot(V)$. 
In the above, all tensor products and direct sums are algebraic -- we do not complete in any topology --  and by convention
$u^{\odot}=1\in V^{\odot 0}=\CC$, $f^{\otimes 0} = \id_\CC$. 
The product may be summarised via the Weyl relations $W(\lambda u)W(\mu v)= e^{-i\lambda\mu
\sigma(u,v)/2} W(\lambda u + \mu v)$, understood as relations between the formal power series
\[
W(\lambda u) = \bigoplus_{n=0}^\infty \frac{(i\lambda)^n }{n!}u^{\odot n}, \qquad (u\in V,~\lambda\in\RR).
\]
(Here, it is not necessary to demand that $u$ is `real', in the sense that $u=Cu$.)

In addition, given any morphism $f:(V,\sigma,C)\to (V',\sigma',C')$ in $\preSympl_\CC$, we define
\begin{equation}\label{eq:Q3}
\Qf(f) = \Gamma_\odot(f) = \bigoplus_{n\in\NN_0}^\infty f^{\odot n}.
\end{equation}

\begin{Prop}\label{prop:Q}
Equations~\eqref{eq:Q1}, \eqref{eq:Q2} and \eqref{eq:Q3} define a functor $\Qf:\preSympl_\CC\to\Alg$. 
If $(V,\sigma,C)$ is weakly nondegenerate, then $\Qf(V,\sigma,C)$ is simple.
\end{Prop}
The proof of this result is largely a matter of assembling standard
results -- it will be given below for completeness.
Before that, we make a number of remarks. 

For obvious reasons, we refer to $\Qf$ as a quantization functor.
Given any classical theory $\Sol:\Mand\to\preSympl_\CC$, 
we obtain a quantum theory $\Af=\Qf\circ\Sol$; if $\Sol$ obeys the timeslice property, then
so does $\Af$, because functors preserve isomorphisms,
and its relative Cauchy evolution is given by 
\begin{equation}
\rce_\Mb[\hb] = \Qf(\rce^{(\Sol)}_\Mb[\hb])  =\bigoplus_{n\in\NN_0} R_\Mb[\hb]^{\otimes n}.
\label{eq:rce_for_Af}
\end{equation}
where, to unburden the notation, we have written 
$\rce$ for $\rce^{(\Af)}$ and $R$ for $\rce^{(\Sol)}$.
Furthermore,  
$\Qf$ interacts well with the unions in $\preSympl_\CC$ and $\Alg$: 
given a (possibly infinite) family of $C$-invariant subspaces of $V$, $W_i$, then
\[
\bigvee_i \Gamma_\odot(W_i) = \Gamma_\odot(\bigvee_i W_i),
\]
where the union on the left-hand side is an algebraic span (in $\Qf(V,\sigma,C)$),
while that on the right is a vector space span in $V$. Inclusion of the left-hand side in the right-hand side
is obvious; the reverse inclusion arises from the freedom to form products as well as linear combinations
in $\Alg$.

We now turn to the proof of Prop.~\ref{prop:Q}, beginning by
giving a construction of $\Qf(V,\sigma,C)$ that allows its various properties to be established. 
Let $\iVect$ be the category whose objects are pairs $(V,C)$, where $C$ is an 
antilinear involution on complex vector space $V$, and with morphisms $f:(V,C)\to
(V',C')$ which are injective linear maps such that $C'\circ f = f\circ C$. 
Then there is a functor $\Tf:\iVect\to\Alg$ which constructs the tensor algebra over given vector spaces:
\[
\Tf(V) = \bigoplus_{n\in\NN_0} V^{\otimes n} ,\qquad \Tf(f) =  \bigoplus_{n\in\NN_0} f^{\otimes n}
\]
where the product in $\Tf(V)$ is given by the tensor product and the $*$-operation by 
\[
(\phi_1\otimes\cdots\otimes\phi_n)^* = C(\phi_n)\otimes\cdots\otimes 
C(\phi_1)
\]
(injectivity of $\Tf(f)$ follows from injectivity of $f$ after some multilinear algebra --- see Appendix~\ref{appx:multilin} for details). As above, all tensor products and direct sums are algebraic. 

There is an obvious forgetful functor from $\preSympl_\CC$ to $\iVect$, and so
$\Tf$ (we use the same notation for $\Tf$ and its composition with the forgetful functor)
can be defined from $\preSympl_\CC\to\Alg$. Given any $(V,\sigma,C)$, let $\Zf(V,\sigma,C)$
be the two-sided $*$-ideal in $\Tf(V,\sigma,C)$ generated by elements of form
\[
(-i\sigma(u,v),0,u\otimes v-v\otimes u,0,\ldots) \qquad (u,v\in V)
\]
and write $\Qf(V,\sigma,C)$ for the quotient $\Tf(V,\sigma,C)/\Zf(V,\sigma,C)$.
If $f:(V,\sigma,C)\to (V',\sigma',C')$ then  $\Tf(f)$ maps $\Zf(V,\sigma,C)$ into
$\Zf(V',\sigma',C')$, which induces a unital $*$-homomorphism $\Qf(f):\Qf(V,\sigma,C)\to \Qf(V',\sigma',C')$. 

As it is clear that $\Zf(V,\sigma,C)$ has trivial intersection with the symmetric subspace of $\Tf(V,\sigma,C)$, it follows that (i) the quotient algebras are nontrivial; (ii) every element $A\in\Tf(V,\sigma,C)$ has
a unique {\em symmetric representative} $A_\odot$, in the symmetric subspace and that $A=0$ iff $A_\odot=0$; (iii) we may therefore identify $\Qf(V,\sigma,C)$ {\em
as a vector space} with $\Gamma_\odot(V)$ (as in~\eqref{eq:Q1})
and any morphism $\Qf(f)$ ($f:(V,\sigma,C)\to (V',\sigma',C')$) may be identified
{\em as a linear map} with the restriction of $\Tf(f)$ to this subspace (as in~\eqref{eq:Q3}); 
(iv) the homomorphisms $\Qf(f)$ are therefore injective and hence
define $\Alg$-morphisms. It is clear that $\Qf$ inherits functoriality from $\Tf$. 

To complete the proof of Prop.~\ref{prop:Q}, we need to verify the product formula
\eqref{eq:Q2} and show that $\Qf(V,\sigma,C)$ is simple when $(V,\sigma,C)$ is weakly nondegenerate. 
The latter follows from Scholium~7.1 in~\cite{BaezSegalZhou}, while the former requires a tedious calculation with commutators if written explicitly. However, 
the mere existence of such an argument indicates that the question is purely one of combinatorics,
and can be resolved by examining the case in which $V$ is of dimension $2$, $\sigma$ is
nondegenerate, $u=\lambda e_1$, $v=\mu e_2$, for $\lambda,\mu\in\RR$, where $\sigma(e_1,e_2)=1$
and $Ce_i=e_i$. Invoking a Fock representation (e.g., with respect
to the Hilbert space norm in which the $e_i$ are orthonormal) the Weyl operators may be
obtained as convergent power series on a domain of analytic vectors and the required
product may be read off as a consequence of the Weyl relations.

To conclude this discussion, we note that our construction is equivalent to a more familiar
quantization of the scalar field (and similar Bose free fields). Let $\Sol_\CC$ be the 
complex Klein--Gordon theory (with mass $m\ge 0$), with corresponding
quantum field theory $\Af=\Qf\circ\Sol_\CC$. Now, for each $f\in\CoinX{\Mb}$ let
\[
\Phi_\Mb(f)= (0,E_\Mb f,0,\ldots) \in \Af(\Mb) = \Gamma_{\odot}(\Sol_\CC(\Mb)).
\]
As $\Sol_\CC(\Mb)$ is precisely the range of $E_\Mb$ on $\CoinX{\Mb}$, it is easy to see that
the $\Phi_\Mb(f)$ generate $\Af(\Mb)$ and obey the relations:
\begin{itemize}\addtolength{\itemsep}{-0.5\baselineskip}
\item $f\mapsto\Phi_\Mb(f)$ is complex linear
\item $\Phi_\Mb(f)^*=\Phi_\Mb(\overline{f})$
\item $\Phi_\Mb(P_\Mb f) = 0$
\item $[\Phi_\Mb(f),\Phi_\Mb(f')]=iE_\Mb(f,f')\II$
\end{itemize}
for all $f,f'\in\CoinX{\Mb}$. In fact, owing to simplicity of $\Af(\Mb)$, it may be completely characterized by these generators and relations. Moreover, for any morphism $\psi:\Mb\to\Nb$, we have
\[
\Af(\psi)\Phi_\Mb(f) = (0, \Sol_\CC(\psi) E_\Mb f,0,\ldots) = 
(0, E_\Nb\psi_*  f,0,\ldots) = \Phi_\Nb(\psi_*f).
\]
In this sense, $\Phi$ may be regarded as a natural transformation between the functor $\Df:\Loc\to\Vect$, 
with $\Df(\Mb)=\CoinX{\Mb}$, $\Df(\psi)=\psi_*$, and the functor $\Af$, after the latter is composed with
a forgetful functor to the category of vector spaces. 
This is the understanding of `quantum fields as natural transformations' 
first articulated by BFV. 

The quantized massless current, $\Cf = \Qf\circ\Jf$, may be treated in the same way. To each $\omega\in\Omega_0^1(\Mb)$, we assign  $\Js_\Mb(\omega)=(0,[E_\Mb\delta_\Mb\omega],0,\ldots)\in \Gamma_\odot(\Jf(\Mb))=
\Cf(\Mb)$.
The surjectivity result in Lemma~\ref{lem:current} establishes that these elements generate $\Cf(\Mb)$; moreover, they 
clearly satisfy the relations
\begin{itemize}\addtolength{\itemsep}{-0.5\baselineskip}
\item $\Omega_0^1(\Mb)\owns \omega\mapsto \Js_\Mb(\omega)$ is complex linear
\item $\Js_\Mb(\omega)^*=\Js_\Mb(\overline{\omega})$, for all $\omega\in\Omega_0^1(\Mb)$
\item $\Js_\Mb(\omega) = 0$ for all $\omega\in \Omega_0^1(\Mb)\cap \ker\delta_\Mb +\Omega_0^1(\Mb)\cap d_\Mb C^\infty(\Mb)$
\item $[\Js_\Mb(\omega),\Js_\Mb(\omega')]=iE_\Mb(\delta_\Mb \omega,\delta_\Mb\omega')\II$ for all 
$\omega,\omega'\in \Omega_0^1(\Mb)$. 
\end{itemize}
The last of these holds because
\[
\sigma_{0\,\Mb}([E_\Mb\delta_\Mb\omega],[E_\Mb\delta_\Mb\omega'])
= \sigma_\Mb(E_\Mb\delta_\Mb\omega,E_\Mb\delta_\Mb\omega')
=E_\Mb(\delta_\Mb\omega,\delta_\Mb\omega').
\]
Moreover, under any $\psi:\Mb\to\Nb$ in $\Loc$, we have the naturality property
$\Cf(\psi)\Js_\Mb(\omega) = \Js_\Nb(\psi_*\omega)$ by a similar
calculation to that used in the standard scalar field case. Although it
is not immediately obvious, the third relation above is compatible with
this map, because $\Cf$ is known to be a functor. A direct proof can be given,
but  will not be done here.

\subsection{Dynamical locality}

Suppose that $\Sol:\Mand\to\preSympl_\CC$ is a weakly nondegenerate locally covariant theory, 
obeying the general conditions ($\Sol1$--$\Sol4$) stated in Sect.~\ref{sect:Klein_Gordon}, but which is not necessarily dynamically local.\footnote{In passing, however, we remark that for theories obeying dynamical locality, ($\Sol3$) is equivalent to 
the absence of nonzero elements invariant under arbitrary relative Cauchy evolution --- see Thm~6.5 of~\cite{FewVer:dynloc_theory}. Moreover, in $(\Sol2)$, conservation of the stress-energy tensor follows from the arguments given in BFV in the quantum case.} As shown above, all these assumptions hold for the massive minimally coupled field and the massless current
and are expected to hold for the dynamically local linear Bose fields of interest. At any rate, as
we will see, these statements isolate the  properties that are necessary to establish dynamical locality of the quantized theory $\Af=\Qf\circ \Sol$ in the case that $\Sol$ is dynamically local.

To begin the discussion, we note that the kinematic
net is easily obtained in terms of that of $\Sol$: for any nonempty $O\in\OO(\Mb)$ we have
$\Af(\iota_{\Mb;O}) = \Qf(\Sol(\iota_{\Mb;O}))$ and hence 
$\Af^\kin(\Mb;O)$ may be identified as the linear subspace
\[
\Af^\kin(\Mb;O) = \Gamma_\odot(\Sol^\kin(\Mb;O))\subset \Af(\Mb).
\]
One might think that there would be a similar `abstract nonsense' computation of the
dynamical nets, given the close relation between the relative Cauchy evolutions of $\Sol$ and $\Af$.
It is certainly true that if $\Qf$ were to preserve equalizers, intersections and unions,
then the dynamical net could be computed in this way. However $\Qf$
does not preserve equalizers\footnote{As an example, consider $(V,\sigma,C)$, where $V=\CC^2$ with a basis $v_i$ ($i=1,2$) obeying $\sigma(v_1,v_2)= 1$ and  $Cv_i=v_i$. Then 
the map $f (a v_1+b v_2) = av_2- bv_1$ ($a,b\in\CC$) defines an automorphism of $(V,\sigma,C)$
whose equalizer with the identity is trivial. However, $v_1\otimes v_1+
v_2\otimes v_2$ is a (nonzero) eigenvector of eigenvalue $1$ for $\Qf(f)$, which therefore has
nontrivial equalizer with the identity in $\Qf(V,\sigma,C)$.} and we must calculate the dynamical subalgebras directly. 

To this end, note that each element $A\in\Af(\Mb)$ may be associated with a finite-dimensional subspace $Y_A$ of $\Sol(\Mb)$ in the following way. 
For each $n\ge 1$,  the component $A_n$ of $A$ in $\Sol(\Mb)^{\odot n}$ may be regarded 
as a linear map $A_n:(\Sol(\Mb)^{\otimes (n-1)})^*\to \Sol(\Mb)$ with finite dimensional image, which we call the {\em support subspace} of $A_n$; 
the support space $Y_A$ of $A$ is defined to be the span of these images for $n\ge 1$, only finitely many of which are nontrivial. (Here, all duals are algebraic.) Moreover, $A\in \Gamma_\odot(Y_A)$. 
Some basic properties of support spaces are summarised in Appendix~\ref{appx:multilin}. The crucial observation is:
\begin{Lem} \label{lem:support_space}
If $A\in\Af^\bullet(\Mb;K)$ then the support subspace $Y_A$ is invariant under $R_\Mb[\hb]$ for all $\hb\in H(\Mb;K^\perp)$ and hence under $F_\Mb[\fb]$ for all $\fb\in \Sym(\Mb;K^\perp)$.
Hence $\Af^\bullet(\Mb;K)\subset \Gamma_\odot(\Sol^\bullet(\Mb;K))$. 
\end{Lem}
{\noindent\em Proof:} As $\rce_\Mb[\hb]A=A$ for all $\hb\in H(\Mb;K^\perp)$, we have $\Gamma_\odot(R_\Mb[\hb]) A =A$
for all such $\hb$, and because the action is diagonal with respect to the direct sum structure, it is clear that the component $A_n$ of $A$ in $\Sol(\Mb)^{\odot n}$ obeys $R_\Mb[\hb]^{\otimes n}A_n = A_n$ for all such $\hb$. By Lem.~\ref{lem:inv_supp}, this entails that the support space $Y_n$ of
$A_n$ is invariant under $R_\Mb[\hb]$ for any $n\ge 1$, so the support space 
$Y_A=\bigvee_{n=1}^\infty Y_n$ is also invariant. The statement regarding invariance under $F_\Mb$ follows, given their definition as functional derivatives of the $R_\Mb$ with respect to the metric,
and we deduce that $Y_A\subset \Sol^\bullet(\Mb;K)$ by Prop.~\ref{prop:inv_to_bullet}. 
Hence $A\in\Gamma_\odot(Y_A)\subset \Gamma_\odot(\Sol^\bullet(\Mb;K))$. 
$\square$

Our main result of this section is that the dynamical nets of $\Af$ are, after all, related to those
of $\Sol$ by the quantization functor $\Qf$. We use the notation $\alpha^{\bullet/\dyn/\kin}_{\Mb;X}$
for the inclusions of $\Af^{\bullet/\dyn/\kin}(\Mb;X)$ in $\Af(\Mb)$, and $\lambda^{\bullet/\dyn/\kin}_{\Mb;X}$ for the corresponding morphisms in the theory $\Sol$. 

\begin{Thm} 
Let $\Sol:\Mand\to\preSympl_\CC$ be any weakly nondegenerate theory obeying assumptions ($\Sol1$)--($\Sol4$) and let $\Af=\Qf\circ\Sol$. 
For any compact set $K$ in $\Mb\in\Mand$, 
\begin{equation}\label{eq:QFT_bullet}
\Af^\bullet(\Mb;K) = \Gamma_\odot(\Sol^\bullet(\Mb;K)),
\end{equation}
as vector spaces, i.e., $\alpha^\bullet_{\Mb;K}\cong \Qf(\lambda^\bullet_{\Mb;K})$. 
For any nonempty $O\in\OO(\Mb)$, 
\begin{equation}\label{eq:QFT_dyn}
\Af^\dyn(\Mb;O) = \Gamma_\odot(\Sol^\dyn(\Mb;O)),
\end{equation}
i.e., $\alpha^\dyn_{\Mb;K}\cong \Qf(\lambda^\dyn_{\Mb;K})$. Hence if, additionally, $\Sol$ is
dynamically local, then $\Af$ is dynamically local. 
\end{Thm}
{\noindent\em Proof.} The inclusion of the left-hand side of \eqref{eq:QFT_bullet} in the
right-hand side is established in Lem.~\ref{lem:support_space}. The reverse inclusion
is immediate from \eqref{eq:rce_for_Af} and the definition of $\Sol^\bullet(\Mb;K)$. 
The second part follows from the fact that $\Qf$ preserves unions.
$\square$

Our application is then immediate. 
\begin{Cor} \label{cor:main1}
The infinitesimal Weyl-algebra quantizations of (a) the Klein--Gordon theory for $m>0$ in any dimension $n\ge 2$,  as a theory on $\Mand$ or $\Man$, and (b) the theory of the massless current in any dimension
$n\ge 2$ as a theory on $\Man$, or $n\ge 3$ as a theory on $\Mand$, are dynamically local. 
The infinitesimal Weyl-algebra quantizations of (c) the massless  Klein--Gordon theory 
in any dimension $n\ge2$, and as a theory on either $\Man$ or $\Mand$, and (d) 
the massless current in dimension $n=2$ as a theory on $\Mand$, are not dynamically local. 
\end{Cor}
For example, in the massless scalar field, we have  
\[
\Af^\dyn(\Mb;O) = \Af^\kin(\Mb;O)\vee \Gamma_\odot(\Sol_{l.c.}(\Mb))
\]
which differs from $\Af^\kin(\Mb;O)$ in any spacetime with (at least one connected component having)
a compact Cauchy surface. The failure of dynamical locality in this case is  related to another pathology of 
the model, namely, the nonexistence of ground states in ultrastatic spacetimes with 
compact Cauchy surface. Again it is the locally constant solutions (usually regarded in terms of the 
zero modes of the spatial Laplacian) that create the problem, which is absent in the theory of 
the massless current. We observe that the rigid $\phi\mapsto \phi+\textrm{const}$ symmetry is spontaneously broken --- it cannot be unitarily implemented even
in the Minkowski vacuum state although it is an automorphism of the algebra~\cite{Streater_broken}; 
at a higher level the symmetry generates automorphisms of the functor $\Af$ \cite{Fewster:gauge}.

\section{Quantized theory: Weyl algebra}\label{sect:Weyl}

In this section, we study the other main approach to the quantization of linear field theories, namely the Weyl algebra approach. We show that dynamically local theories valued in $\preSympl_\RR$ have
dynamically local quantizations under mild additional conditions: essentially it is enough that
the symplectic products of the classical theory are nondegenerate (so the theory factors via
$\Sympl_\RR$) and the relative Cauchy evolution acts continuously in a certain sense. These conditions
are met by the massive Klein--Gordon theory and the theory of currents. It seems reasonable to 
expect that the Proca and (with some care) electromagnetic fields could also be shown dynamically local. 
See~\cite{Few&Pfen03} for the algebraic approach to the quantization of the Proca field and~\cite{Dapp:2011} for details on its locally covariant formulation; for electromagnetism, 
see~\cite{Dimock92, Few&Pfen03, Dapp:2011} for the formulation in terms of vector potentials
and~\cite{DappLang:2011} and Appendix~A of~\cite{Hollands:2008} treatments in terms of field strength. 

We begin by recalling some relevant background. If $(S,\sigma)\in\Sympl_\RR$, the category of weakly nondegenerate real symplectic spaces\footnote{See~\cite{BiHoRi2004} for the construction of the Weyl algebra over pre-symplectic spaces.} the CCR algebra 
$\CCR(S,\sigma)$ may be defined as $C^*$-subalgebra of the bounded linear operators
on $\ell^2(S)$ generated by operators
$\{W(u):u\in S\}$ with action
$(W(u)f)(v) = e^{i\sigma(u,v)/2} f(u+v)$ ($f\in\ell^2(S)$). These operators evidently
obey the Weyl relations
\[
W(0)= 1, \qquad W(u)^*=W(-u), \qquad W(u+v) = e^{i\sigma(u,v)/2} W(u)W(v).
\]
We write $\WW(S,\sigma)$ for the $*$-algebraic span of the $W(u)$ (which simply amounts to their linear span given the Weyl relations). 
Accordingly, any $A\in \CCR(S,\sigma)$ may be written as a limit (in operator norm) 
\[
A = \lim_{n\to\infty} \sum_{u\in S} a_n(u) W(u)
\]
where, for each $n$, at most finitely many $a_n(u)$ are nonzero, and there are at most
countably many $u$ for which there is any $n$ with $a_n(u)\neq 0$.\footnote{The point is that
$A$ can be realised as a limit of a sequence in $\WW(S,\sigma)$, each term of which
involves only finitely many $u\in S$. Thus at most countably many such $u$ appear in 
this sequence as a whole.} Considering matrix elements, it is clear that we have limits
$a(u)=\lim_n a_n(u)$ for each $u$, at most countably many of which are nonzero. 
Let $e_u$ be the basis vector in $\ell^2(S)$ labelled by $u\in S$: $e_u(v)=\delta_{u v}$. 
Then
\[
Ae_0= \lim_n \sum_{u\in S} a_n(-u) e_u = \sum_{u\in S} a(-u) e_u
\]
and $\sum_{u\in S} |a(u)|^2=\|Ae_0\|^2<\infty$. 

As is well-known, $\CCR$ is a functor from $\Sympl_\RR$ to $\CAlg$ (see, e.g., 
BFV or Sec. 4.2 of~\cite{BarGinouxPfaffle}). Accordingly, if $\Sol$ is any 
weakly nondegenerate theory $\Sol:\Mand\to\preSympl_\RR$, then we obtain the
Weyl-algebra quantization of this theory as the functor $\Wf:\Mand\to\CAlg$ given by $\Wf=\CCR\circ\Sol$ (slightly abusing notation by regarding $\Sol$ as being valued in $\Sympl_\RR$, rather than $\preSympl_\RR$). 

We will need some information concerning fixed-points. If $\alpha$ is an automorphism of $\CCR(S,\sigma)$ induced by symplectic automorphism $g$
of $(S,\sigma)$ then 
\[
\alpha A = \lim_{n\to\infty} \sum_{u\in S} a_n(u) W(g u) =  
\lim_{n\to\infty} \sum_{u\in S} a_n(g^{-1} u) W(u) .
\]
Thus if $\alpha A = A$, we have in particular (by considering the action on $e_0$) that
$a(u) = a(g^{-1}u)$ for all $u$. As $\sum_{u\in S} |a(u)|^2$ is finite, it follows that
$a(u)$ can be nonzero only for $u$ such that $g^k u=u$ for some $k\in\ZZ$. Thus the fixed-point subalgebra $\CCR(S,\sigma)^\alpha$ is the closed linear span (in $B(\ell^2(S))$) of elements of the form $\sum_{j=0}^{k-1} W(g^j u)$ for $g^k u=u$. 

More generally, if we consider
the fixed-point subalgebra relative to a group of automorphisms $G$ of $(S,\sigma)$
we must restrict to elements of form $\sum_{v\in Gu} W(v)$ for $u$ such that the orbit $Gu$ of $u$ under $G$ is finite. 

\begin{Prop} \label{prop:CCR_fix}
Let $I\subset \RR$ be an open interval containing the origin, and let $(g(s))_{s\in I}$
be a family of automorphisms of $(S,\sigma)$, with $g(0)=\id_{(S,\sigma)}$, generating a subgroup $G\subset \Aut(S,\sigma)$. 
If there is any Hausdorff topology on $S$ for which $s\mapsto g(s)$ acts continuously on $S$, 
then 
\[
\CCR(S,\sigma)^G = \cl\Span\{W(u): u\in S,~g(s)u=u~\forall s\in I\}.
\]
\end{Prop}
{\noindent\em Proof:} The inclusion of the right-hand side in the left is obvious. 
On the other hand, the remarks above show imply that $\CCR(S,\sigma)^G$ is generated
by (at most) those $W(u)$ for which $\{g(s)u: s\in I\}$ is a finite set. But as there
is a Hausdorff topology on $S$ so that $s\mapsto g(s)u$ is continuous, it follows that 
$g(s) u$ is constant and therefore equal to $g(0)u=u$ for all $s\in I$. $\square$

\begin{Thm} 
Let $\Sol:\Mand\to\preSympl_\RR$ be any weakly nondegenerate dynamically local theory. Suppose that for each $\fb\in\Sym(\Mb)$, $s\mapsto \rce^{(\Sol)}_\Mb[s\fb]$ 
acts continuously in the weak symplectic topology on some open neighbourhood of $s=0$. 
Then $\Wf =\CCR\circ\Sol:\Mand\to\CAlg$ is dynamically local. 
\end{Thm}
{\noindent\em Proof:} Let $\Mb\in \Mand$ be arbitrary, and $O\in\OO(\Mb)$ be nonempty. We first observe that $\Wf^\kin(\Mb;O)$ is
generated by Weyl generators $W_\Mb(u)$ indexed by $u\in\Sol^\kin(\Mb;O)$. 
Next,  the weak symplectic topology separates points on $(\Sol(\Mb),\sigma_\Mb)$, because $\sigma_\Mb$ is weakly nondegenerate, and therefore defines a Hausdorff locally convex topology on $S$.\footnote{We follow the definitions of~\cite[Ch.~V]{ReedSimon:vol1}.}  
By Prop.~\ref{prop:CCR_fix}, 
\[
\Wf^\bullet(\Mb;K) =  \cl\Span\{W(u): u\in \Sol^\bullet(\Mb;K)\}  \subset 
\Wf^\kin(\Mb;O)
\]
because $\Sol^\bullet(\Mb;K)\subset \Sol^\kin(\Mb;O)$ by dynamical locality of $\Sol$. Taking
the $C^*$-union over $K\in\KK_b(\Mb;O)$, we find $\Wf^\dyn(\Mb;O)\subset \Wf^\kin(\Mb;O)$. 

On the other hand, let $u\in \Sol^\kin(\Mb;O)$. Using dynamical locality of $\Sol$, we have
\[
u\in \Sol^\dyn(\Mb;O) = \bigvee_{K\in\KK_b(\Mb;O)} \Sol^\bullet(\Mb;K)
\]
so $u$ may be expressed as a finite sum $u=\sum_i u_i$, with $u_i\in \Sol^\bullet(\Mb;K_i)$ and
with $K_i\in \KK_b(\Mb;O)$. Hence the Weyl generator $W_\Mb(u)$
may be expressed as a finite product of elements in the subalgebras $\Wf^\bullet(\Mb;K_i)$
and is, in particular, contained in $\Wf^\dyn(\Mb;O)$. As $\Wf^\kin(\Mb;O)$ is
generated by Weyl generators of this type, we have $\Wf^\kin(\Mb;O)\subset\Wf^\dyn(\Mb;O)$,
which concludes the proof of dynamical locality for $\Wf$. 
$\square$ 

\begin{Cor} \label{cor:main2}
The Weyl algebra quantizations of (a) the Klein--Gordon theory for $m>0$ in any dimension $n\ge 2$,  as a theory on $\Mand$ or $\Man$, and (b) the theory of the massless current in any dimension
$n\ge 2$ as a theory on $\Man$, or $n\ge 3$ as a theory on $\Mand$, are dynamically local. 
The Weyl algebra quantizations of (c) the massless  Klein--Gordon theory 
in any dimension $n\ge2$, and as a theory on either $\Man$ or $\Mand$, and (d) 
the massless current in dimension $n=2$ as a theory on $\Mand$, are not dynamically local. 
\end{Cor}
{\noindent\em Proof:} We need only add that in the case of the massless field, the arguments
above show that $\Wf^\dyn(\Mb;O)$ contains $\Wf^\kin(\Mb;O)\vee \Wf_{l.c.}(\Mb)$,
where $\Wf_{l.c.}(\Mb)$ is the subalgebra of $\Wf(\Mb)$ generated by solutions in $\Sol_{l.c.}(\Mb)$.
$\square$

\section{Conclusion}\label{sect:outlook}

We have shown that the class of dynamically local theories contains at least the basic examples of
free quantum fields, with the caveat that the massless Klein--Gordon field should be formulated
as the theory of the massless current. Our results were obtained by first showing that our theories
of interest are dynamically local as classical symplectic theories and then by applying general theorems which lift  dynamical locality to the quantized theory. The existence of these theorems illustrates again the natural nature of the dynamical locality definition.

We should like to make some comments about related work. In \cite{BruGuiLo:2002}, the authors start from an
(anti-)unitary representation $u$ of the proper Poincar\'e group
in $1+d$ dimensions on a Hilbert space $\mathcal{H}$ (later
interpreted as a ``one-particle space''). They consider the wedge-region
$\mathcal{W} = \{ (x^0,x^1,\ldots,x^d) : x^1 > 0, \ -x^1 <x^0<x^1 \}$ in
$1+d$-dimensional Minkowski spacetime, and the associated wedge-reflection
symmetry $r_{\mathcal{W}} : (x^0,x^1,x^2,\ldots,x^d) \mapsto (-x^0,-x^1,x^2,\ldots,x^d)$
as well as the one-parametric group $\Lambda_{\mathcal{W}}(t)$ $(t \in \mathbb{R})$
of Lorentz boosts leaving the region $\mathcal{W}$ invariant. Setting
$\delta_{\mathcal{W}}^{1/2} = u(\Lambda_{\mathcal{W}}(i))$ and $j_\mathcal{W} =
u(r_{\mathcal{W}})$, they define a ``one-particle Tomita operator''
$s_{\mathcal{W}} = j_{\mathcal{W}} \delta_{\mathcal{W}}^{1/2}$, and a one-particle
subspace
\[
\mathcal{K}_{\mathcal{W}} = \{ \xi \in {\rm dom}(\delta_{\mathcal{W}}^{1/2}) : 
 s_{\mathcal{W}}\xi = \xi \}\subset\mathcal{H}.
\]
This is regarded as the one-particle subspace of $\mathcal{H}$
consisting of one-particle wave functions localized in $\mathcal{W}$, in the spirit of a
``reverse interpretation'' of the Bisognano-Wichmann theorem \cite{Haag}.
Passing to the second quantization on the Fock space $\mathcal{F}_\odot(\mathcal{H})$,
they associate the von Neumann algebras $\mathcal{R}(\mathcal{W}) =
\{ W(\xi) : \xi \in \mathcal{K}_{\mathcal{W}}  \}''$ with the spaces $\mathcal{K}_\mathcal{W}$,
where $W(\xi)$ is the Weyl-operator of $\xi \in \mathcal{H}$ on $\mathcal{F}_\odot(\mathcal{H})$. 
By forming
local von Neumann algebras of observables
\[
\mathcal{R}(\mathcal{O}) = \bigcap_{L\mathcal{W} \supset \mathcal{O}} 
 \mathcal{R}(L \mathcal{W}) 
\]
where $\mathcal{O}$ is a double cone, and $L \mathcal{W}$ is the image of
$\mathcal{W}$ under any Poincar\'e transform $L$, one obtains a net of local 
observable algebras complying with the Haag-Kastler axioms, under mild, generic
additional assumptions (most importantly, positivity of the energy).

The definition of $\mathcal{K}_\mathcal{W}$---inspired by the  Bisognano-Wichmann theorem---is, in some ways,
analogous to our requirement of invariance under suitable relative Cauchy evolutions. The analogy becomes
somewhat more obvious on noting that $\mathcal{K}_\mathcal{W}$ could 
equivalently be characterized as the symplectic complement, in
$\mathcal{H}$, of $\mathcal{K}_{\mathcal{W}'}$, 
where $\mathcal{W}' = r_{\mathcal{W}}(\mathcal{W})$ is the causal complement wedge of $\mathcal{W}$
(cf.~\cite[Thm~2.5]{BruGuiLo:2002}). However, the analogy does not seem,
as yet, to carry much further, since the elements of the Poincar\'e group 
(or, equivalently, the Tomita-Takesaki modular objects) act globally, whereas
the relative Cauchy evolution in our setting acts locally. There is clearly room for further
investigation of potential relations beyond this analogy.

To conclude, we note that the strategy developed here could be applied to other linear Bose theories (and, with modifications, to linear Fermi theories as well). For example, given 
any collection of weakly nondegenerate dynamically local theories $\Sol_i:\Loc\to\preSympl_\CC$ obeying ($\Sol1$--$\Sol4$), we may form the algebraic direct sum theory $\Sol$, with
\[
\Sol(\Mb) = \bigoplus_i \Sol_i(\Mb),\qquad \Sol(\psi) = \bigoplus_i \Sol_i(\psi)
\]
for any $\Mb\in\Loc$, and $\Loc$ morphism $\psi$. As we work with algebraic direct sums, there is no
issue concerning the convergence of symplectic products etc, even when $i$ runs over an infinite index set. This is clearly a weakly nondegenerate functor to $\preSympl_\CC$ obeying ($\Sol1$--$\Sol4$) because of the direct sum structure. Moreover, it is equally clear that this new theory 
is dynamically local.  Then the theory $\Qf\circ\Sol$ is dynamically local; similarly, we would have this for the Weyl algebra theory $\CCR\circ\Sol$, provided that the relative Cauchy evolution acts continuously
in the weak symplectic topology for each $\Sol_i$. This establishes the dynamical locality of arbitrary multi-component minimally coupled scalar fields, with arbitrary mass spectrum [treating any zero mass components using massless current theory, and subject to the same constraints
on the spacetime dimension as in Corollaries~\ref{cor:main1} and~\ref{cor:main2}]. 

The failure of dynamical locality for the massless current in two-dimensional
spacetimes with disconnected components bears some analogy to the
occurrence of topological superselection sectors in the short-distance scaling
limit of the massive free scalar field on two-dimensional Minkowski spacetime
discussed in the scaling algebra framework in \cite{BucVerScalAlg2}. Interestingly,
there is a dynamical constraint involved in the construction of scaling algebra and
scaling limit, and it appears that there might be a deeper connection between
dynamical locality and the occurrence of topological charges, a point worthy of
further investigation.

In summary, we have established that the class of dynamically local theories contains many interesting theories; others will be studied elsewhere. Theories with pure gauge degrees of freedom will not generally be expected to satisfy dynamical locality, without further modification.

\vspace{0.2cm}
{\small\noindent {\bf Acknowledgments} We thank the organisers of the workshop `Rigorous quantum field theory in the LHC era', held at
the Erwin Schr\"odinger Institute, Vienna, 2011, at which this work was completed, and the ESI for financial support. CJF thanks David Hunt, Ko Sanders and Benjamin Lang for useful comments.}

\appendix

\section{Support subspaces}\label{appx:multilin}

In the body of the paper, we made use of some simple observations on linear algebra in tensor products.
Although these are presumably known, we include details for completeness. In the following, 
$X_1$ and $X_2$ are vector spaces of possibly infinite dimension over $\CC$, and $\otimes$ denotes
the standard tensor product of vector spaces. Recall that every bilinear map $T:X_1\times X_2\to Y$ (for $Y$ any vector space) 
induces a unique linear map $\tilde{T}:X_1\otimes X_2\to Y$ so that $\tilde{T}(x_1\otimes x_2) = T(x_1,x_2)$ for all $x_i\in X_i$. 
In particular, this gives linear maps
\[
L(X_1^*,X_2) \xlongleftarrow{\rho_1} X_1\otimes X_2 
\xlongrightarrow{\rho_2} L(X_2^*,X_1)
\]
so that, for example, $\rho_2$ is defined by extension of the bilinear map 
\[
X_1\times X_2\owns (u,v) \mapsto v^{**}(\cdot) u \in L(X_2^*,X_1),
\]
where $v^{**}$ is the canonical embedding of $v\in X_2$ into $X_2^{**}$.
Now, any $\phi\in X_1\otimes X_2$ may be written as a finite sum
$\phi=\sum_i u_i\otimes v_i$
with nonzero $u_i\in X_1$, $v_i\in X_2$, so 
$\rho_2(\phi)(\cdot) = \sum_i v_i^{**}(\cdot)u_i$
is a finite rank map, as is $\rho_1(\phi)$. Moreover, by combining and possibly discarding terms, we may assume that the $u_i$ and $v_i$ in the expansion of $\phi$ each form linearly independent sets, which are then easily seen to span the images of $\rho_1(\phi)$ and $\rho_2(\phi)$ respectively (consider the applications of these maps to dual bases to the bases formed by extensions of the $u_i$ and $v_i$). We see then that $\phi\in \im\rho_1(\phi)\otimes \im\rho_2(\phi)$,
which also proves that the $\rho_i$ are injective. 

More generally, given any vector spaces $X_1,\ldots, X_n$, we have injections
\[
\rho_k: X_1\otimes\cdots\otimes X_n\to L((X_1\otimes \cdots \widehat{X_k}\cdots \otimes X_n)^*, X_k)
\]
where the hat denotes an omitted factor, and any $\phi\in X_1\otimes\cdots\otimes X_n$ obeys
\[
\phi\in \im\rho_1(\phi)\otimes\cdots\otimes \im\rho_n(\phi).
\]
We refer to the subspaces $\im\rho_k(\phi)\subset X_k$ as {\em support subspaces} of $\phi$. If all the spaces $X_1,\ldots X_n$ are the same, and $\phi$ is an element of the symmetric or antisymmetric subspaces of $X^{\otimes n}$, then all the support subspaces are identical. 

\begin{Lem} \label{lem:inv_supp}
Suppose vector spaces $X_1,\ldots,X_n$ and $Y_1,\ldots, Y_n$ are given ($n\ge 1$), with $S_i\in L(X_i,Y_i)$ for 
each $i$, and define $T_n= S_1\otimes \cdots\otimes S_n$. (a) If each $S_i$ is monic then $T_n$ is also injective. (b) If $Y_i=X_i$ for each $i$ and
$\phi\neq 0$ is an eigenvector of $T_n$ with eigenvalue $\lambda\neq 0$, then
for each $1\le k\le n$, $S_k$ restricts to an automorphism of the support subspace $\im\rho_k(\phi)$.
\end{Lem}
{\noindent\em Proof:} (a) We argue by induction on $n$: suppose this is known to be true for some $n\ge 1$ (it is true for
$n=1$). If $T_{n+1}\phi=0$ then
\[
0 = \rho_{n+1}(T_{n+1}\phi) = S_{n+1} \circ \rho_{n+1}(\phi) \circ T_{n}^* .
\]
Now $T_n$ is injective by the inductive hypothesis, so $T_n^*$ is surjective (see e.g.~\cite[\S2.28]{Greub:linalg}), while  
$S_{n+1}$ is injective by hypothesis.  Thus $\rho_{n+1}(\phi)=0$
and thus $\phi=0$ by injectivity of $\rho_{n+1}$. Hence $T_{n+1}$ is injective and the result follows.
(For a different argument, see~\cite[\S1.18]{Greub:multilin}.)

For (b), we have 
\[
S_k\circ\rho_k(\phi)\circ U_k^* = \rho_k(T_n\phi) = \lambda\rho_k(\phi)
\]
for each $k=1,\ldots, n$, where $U_k = S_1\otimes \cdots \widehat{S_k}\cdots \otimes S_n$. As $\lambda\neq 0$,  it follows that $\im\rho_k(\phi)\subset S_k(\im\rho_k(\phi))$, which, on dimension-counting grounds, is possible only if
$S_k$ restricts to an isomorphism of the support space $\im\rho_k(\phi)$ with itself.
$\square$

\section{Differentiability of classical relative Cauchy evolution}\label{appx:diff}

We indicate how differentiability of $\rce_\Mb^{\Sol_\Kbb}$ may be
established in the weak symplectic topology for $\Kbb=\RR,\CC$. By convention, 
we take spaces of smooth functions to be $\Kbb$-valued. Fix a 
relatively compact open set $O\subset \Mb$ and 
Cauchy surfaces $\Sigma_i$ ($1\le i\le 4$) with $\Sigma_{i+1}\subset I_\Mb^+(\Sigma_i)$
for $i=1,2,3$ and $\cl(O)\subset I_\Mb^+(\Sigma_2)\cap I_\Mb^-(\Sigma_3)$. 
In general, we write $\Mb_{i,j}$ for $I_\Mb^+(\Sigma_i)\cap I_\Mb^-(\Sigma_j)$. 
To start with, let $\hb\in H(\Mb;O)$ be any hyperbolic perturbation supported in $O$.

Owing to the time-slice property, it suffices to study the action of $\rce_\Mb[\hb]$ on
$E_\Mb\CoinX{\Mb_{3,4}}$. Fix $\chi\in C^\infty(\Mb)$, with
$\chi=0$ on $I^+_\Mb(\Sigma_2)$ and $\chi=1$ on $I_\Mb^-(\Sigma_1)$. 
Then the identity
\begin{equation}\label{eq:claim1}
\rce_\Mb[\hb] E_\Mb = E_\Mb P_\Mb\chi E_{\Mb[\hb]} = E_\Mb P_\Mb\chi E^-_{\Mb[\hb]} 
\end{equation}
holds on $\CoinX{\Mb_{3,4}}$. Next, the identity $P^{\phantom{-}}_{\Mb}E_{\Mb[\hb]}^- f = f + (P_\Mb-P_{\Mb[\hb]})E_{\Mb[\hb]}^- f$
entails (as both terms on the right-hand side are compactly supported)
\[
E_{\Mb[\hb]}^- f =E_{\Mb}^- f - E_{\Mb}^-K_{\Mb[\hb]} E_{\Mb[\hb]}^- f
\]
by uniqueness of advanced solutions to the inhomogeneous equation,
where $K_{\Mb[\hb]} = P_{\Mb[\hb]}-P_{\Mb}$. Iterating this formula, 
\begin{equation}
E_{\Mb[\hb]}^- f =E_{\Mb}^- f - E_\Mb^- K^{\phantom{-}}_{\Mb[\hb]} E_{\Mb}^- f +
E_\Mb^- K^{\phantom{-}}_{\Mb[\hb]}E_\Mb^- K^{\phantom{-}}_{\Mb[\hb]} E_{\Mb[\hb]}^- f
\end{equation}
for all $f\in\CoinX{\Mb}$. Substituting this in \eqref{eq:claim1}, we have
\[
(\rce_\Mb[\hb] -\id) E_\Mb f  = - E_\Mb P_\Mb\chi E_\Mb^- K_{\Mb[\hb]} (E_\Mb^-f  -  E_\Mb^-K^{\phantom{-}}_{\Mb[\hb]} E_{\Mb[\hb]}^- f)
\]
for $f\in\CoinX{\Mb_{3,4}}$. Now as $\supp\chi$ lies to the past of the support of $K_{\Mb[\hb]} \phi$ for any
smooth $\phi$, we may replace $\chi E_\Mb^-$ by $\chi E_\Mb$ in the above formula, and use the fact that
$E_\Mb P_\Mb\chi E_\Mb = E_\Mb$. Thus
\begin{align*}
(\rce_\Mb[\hb] -\id) E_\Mb f  &= - E_\Mb K_{\Mb[\hb]} (E_\Mb^-f  -  
E_\Mb^-K^{\phantom{-}}_{\Mb[\hb]} E_{\Mb[\hb]}^- f) \\
&=  - E^{\phantom{-}}_\Mb K^{\phantom{-}}_{\Mb[\hb]} E^{\phantom{-}}_\Mb f +  E^{\phantom{-}}_\Mb K^{\phantom{-}}_{\Mb[\hb]}  
E_\Mb^- K^{\phantom{-}}_{\Mb[\hb]} E_{\Mb[\hb]}^- f.
\end{align*}
Taking symplectic products with $\phi'\in \Sol_\Kbb(\Mb)$, we obtain
\[
\sigma_\Mb((\rce_\Mb[\hb] -\id) E_\Mb f,\phi') = 
-\int_\Mb \phi' K_{\Mb[\hb]} E_\Mb f \dvol_\Mb + 
\int_\Mb \phi' K^{\phantom{-}}_{\Mb[\hb]}  
E^{-}_\Mb K^{\phantom{-}}_{\Mb[\hb]} E_{\Mb[\hb]}^- f \dvol_\Mb. 
\]
Now put $\hb = s\fb$ for $\fb\in\Sym(\Mb;O)$. We will argue below that the second integral in the previous
formula is of order $O(s^2)$ as $s\to 0$, with $\phi'$ and $f$ fixed. Accordingly, 
\begin{align*}
\left.\frac{d}{ds} 
\sigma_\Mb(\rce_\Mb[\hb] \phi,\phi') \right|_{s=0} &=
-\lim_{s\to 0} \frac{1}{s} \int_\Mb \phi' K_{\Mb[s\fb]} \phi\dvol_\Mb  \\
&= \int_\Mb \phi' 
\left( 
\frac{1}{2}\left(\nabla^a
f{}^b{}_b\right)\nabla_a\phi
-\nabla_a f^{ab} \nabla_b\phi\right) \dvol_\Mb  \\
&= \sigma_\Mb(F_\Mb[\fb]\phi,\phi'),
\end{align*}
where $F_\Mb[\fb]$ is as defined in the body of the paper. 

To complete the proof we need to establish the $O(s^2)$ behaviour for the integral given above. 
The integrand is supported in $\Mb_{2,4}$, on which both $\phi'$ and
$K^{\phantom{-}}_{\Mb[\hb]}  
E^{-}_\Mb K^{\phantom{-}}_{\Mb[\hb]} E_{\Mb[\hb]}^- f $ are square-integrable.
It is therefore enough to show that the $L^2$-norm of the latter is $O(s^2)$ as $s\to 0$. 
This may be done by means of energy estimates, which show that
\[
\left\|E^-_{\Mb[s\fb]} f \right\|_{p}  \le C_p \|f\|_{p-1}\qquad (f\in \CoinX{\Mb_{2,4}})
\]
for any $p\ge 1$, with $C_p$ uniform in $s$ for sufficiently small $s$,
and where $\|\cdot\|_p$ is an energy norm on the strip $\Mb_{2,4}$. (See, e.g., the proof of Thm~3.7 of~\cite[Appx~3]{ChoquetBruhat:GR},
noting that the condition of Sobolev regularity holds as we are able to work within a finite number of charts.)
Thus we have
\[
\left\|K^{\phantom{-}}_{\Mb[s\fb]}  
E^{-}_\Mb K^{\phantom{-}}_{\Mb[s\fb]} E_{\Mb[\fb]}^- f \right\|_0
\le C_2 C_3 \left\|K^{\phantom{-}}_{\Mb[s\fb]}  \right\|_{0,2} 
\left\|K^{\phantom{-}}_{\Mb[s\fb]}  \right\|_{1,3}
\left\| f \right\|_{2}
\]
and the required estimate follows because the first two norms on the right-hand side are each 
of order $O(s)$ as $s\to 0$.

\end{document}